\documentclass[apj]{emulateapj}

\usepackage{graphicx}
\usepackage{setspace}

\pdfoutput=1

\slugcomment{Accepted for Publication in ApJ January 8th 2016}

\shorttitle{Marginalising instrument systematics in HST WFC3 transit lightcurves}
\shortauthors{Wakeford et al.}

\begin{document}


\title{Marginalising instrument systematics in HST WFC3 transit lightcurves}


\author{H. R. Wakeford\altaffilmark{1,2}}
\affil{NASA Goddard Space Flight Center, Greenbelt, MD 20771, USA}
\email{hannah.wakeford@nasa.gov}

\author{D. K. Sing\altaffilmark{2} and T. Evans\altaffilmark{2}}
\affil{University of Exeter, Exeter, Devon, UK, EX4 4QL}

\author{D. Deming\altaffilmark{3}}
\affil{University of Maryland, College Park, MD 20742, USA}

\and

\author{A. Mandell \altaffilmark{1}}
\affil{\ }




\begin{abstract}
Hubble Space Telescope (HST) Wide Field Camera 3 (WFC3) infrared observations at 1.1-1.7\,$\mu$m probe primarily the H$_2$O absorption band at 1.4\,$\mu$m, and has provided low resolution transmission spectra for a wide range of exoplanets. We present the application of marginalisation based on \citet{gibson2014} to analyse exoplanet transit lightcurves obtained from HST WFC3, to better determine important transit parameters such as R$_p$/R$_*$, important for accurate detections of H$_2$O. 
We approximate the evidence, often referred to as the marginal likelihood, for a grid of systematic models using the Akaike Information Criterion (AIC). We then calculate the evidence-based weight assigned to each systematic model and use the information from all tested models to calculate the final marginalised transit parameters for both the band-integrated, and spectroscopic lightcurves to construct the transmission spectrum. We find that a majority of the highest weight models contain a correction for a linear trend in time, as well as corrections related to HST orbital phase. We additionally test the dependence on the shift in spectral wavelength position over the course of the observations and find that spectroscopic wavelength shifts $\delta_\lambda(\lambda)$, best describe the associated systematic in the spectroscopic lightcurves for most targets, while fast scan rate observations of bright targets require an additional level of processing to produce a robust transmission spectrum. 
The use of marginalisation allows for transparent interpretation and understanding of the instrument and the impact of each systematic evaluated statistically for each dataset, expanding the ability to make true and comprehensive comparisons between exoplanet atmospheres. 
\end{abstract}




%
%
\section{Introduction}
H$_2$O is the most spectroscopically dominant species expected in hot Jupiter atmospheres, and a key molecules for constraining atmospheric compositions. In most lower atmosphere models of hot Jupiters, H$_2$O is well mixed throughout the  atmosphere, and most features between 0.7 and 2.5\,$\mu$m are the result of H$_2$O vibrationan-rotation bands (\citealt{brown2001}). Hubble Space Telescope (HST) Wide Field Camera 3 (WFC3) infrared observations at 1.1-1.7\,$\mu$m probe primarily the H$_2$O absorption band at 1.4\,$\mu$m, and has provided low resolution transmission spectra for a wide range of exoplanets (e.g. \citealt{berta2012}; \citealt{gibson2012b}; \citealt{deming2013}; \citealt{wakeford2013}; \citealt{knutson2014}; \citealt{stevenson2014c}; \citealt{Sing2015}). Using both transmission and emission spectra \citealt{kreidberg2014b} measured the relative H$_2$O abundance in the atmosphere of WASP-43b. They compare the measured H$_2$O abundance with the solar system giant planets using the metallicity expected for a `broadly' solar case, indicating that the trend observed in the metal abundance of the solar system giant planet atmospheres, i.e. decreasing metal enhancement with increasing mass, extends out to exoplanet atmospheres. \citet{wakeford2013} were also able to obtain a strong detection of H$_2$O in the atmosphere of the hot Jupiter HAT-P-1b. The analysis made the use of a common spectral type star in the field of view of the WFC3 detector and performed differential spectrophotmetry using the companion spectrum, similar to methods used from the ground. The robust nature of the transmission spectral shape shown for analysis methods with and without the use of a companion spectrum, shows that the emphasis needs to be placed on methods that not only effectively reduce the data, but ones that can be applied successfully to multiple datasets simultaneously for a true comparative study.

A wide range of reduction and analysis methods have been used on HST WFC3 transit datasets, making comparisons between datasets and planetary atmospheres difficult. Studies of multiple WFC3 datasets have been conducted across multiple programs and observation modes. These have attempted to define a common de-trending technique to apply to all datasets (\citealt{deming2013}; \citealt{mandell2013}; \citealt{ranjan2014}; \citealt{kreidberg2014a}, \citealt{kreidberg2014b}; \citealt{haynes2015}). WFC3 has two main observing modes that are commonly used for transiting exoplanet spectra; stare  and spatial scan mode. Stare mode has been used for a majority of HST observations and is useful when observing dimmer target stars where the photon counts/pixel/second is low; observing brighter targets in this mode leads to saturation. Stare mode maintains a constant pointing of the telescope throughout the observation, maintaining the same pixel position on the detector. Spatial scanning mode was made available on WFC3 in Cycle 19 (2012) and is now implemented as the main mode of observations for transiting exoplanets, as targets observed for atmospheric follow-up with such instruments often orbit brighter target stars (V magnitudes brighter than 11). WFC3 spatial scanning involves nodding the telescope during an exposure to spread the light along the cross-dispersion axis, resulting in a higher number of photons by a factor of ten per exposure while considerably reducing overheads. This also increases the time of saturation of the brightest pixels, which allows for longer exposure times (\citealt{mccullough2011}). 

\citet{mandell2013} conducted the first reanalysis test of WFC3 data for \mbox{WASP-19b} and \mbox{WASP-12b}, with the addition of the first analysis of \mbox{WASP-17b}. The reanalysis incorporated a wavelength dependent systematic correction over the methods used in the previous analysis. While this study produced almost photon-limited results in individual spectral bins, the spectral features observed in the transmission spectra were degenerate with various models of temperature and compositions making interpretation difficult. \citet{ranjan2014} conducted a study of four stare mode WFC3 transit lightcurves from the large HST program, \mbox{GO\,12181}. However, they were unable to resolve any features in the transmission spectrum for three of the planetary atmospheres, and were unable to extract a robust transmission spectrum for one of the datasets, as different treatments to the data gave moderately different results. 

\begin{figure*}
\centering 
	\includegraphics[width=0.95\textwidth]{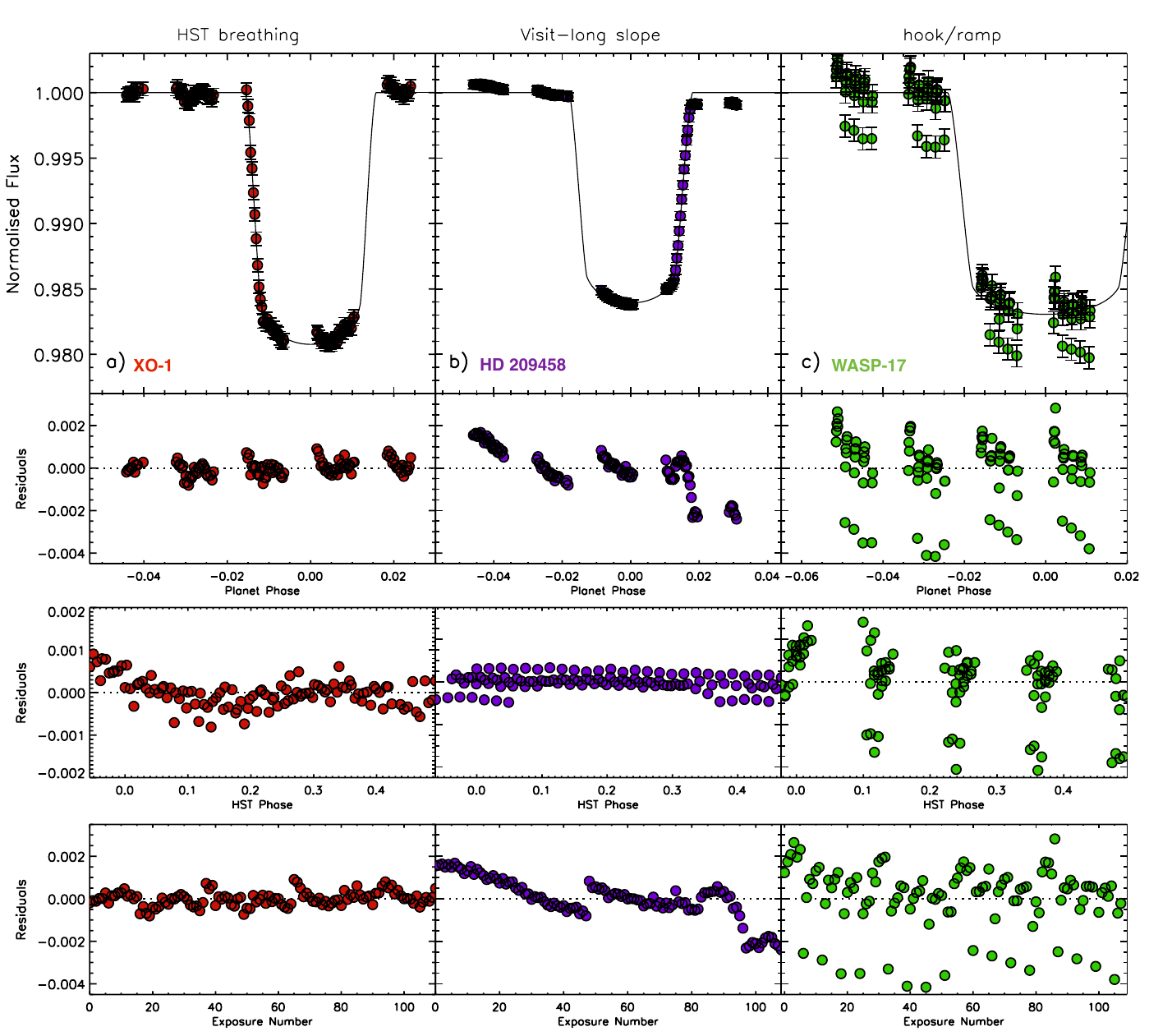}
\caption{Three of the main systematic effects observed in HST WFC3 transit datasets: a) ``HST breathing'' effect caused by the temperature variations in the orbital period of HST. b) Visit-long Slope, a linear trend observed across the entire observing period for all transit lightcurves observed with HST WFC3. c) The `ramp' effect which is thought to be caused by charge trapping between buffer dumps. Lower panels show the residuals of each dataset with respect to different timeseries parameters. Top: residuals in terms of planetary phase. Middle: residuals in terms of HST orbital phase, where each HST orbit of data is overplotted on subsequent orbits. Bottom: residuals in terms of exposure number.}
\label{fig:WFC3_systematics}
\end{figure*}
Additionally, the analysis of the very hot Jupiter \mbox{WASP-12b} which was observed as part of \mbox{GO\,12230}, P.I. M. Swain is a good example of differences caused from analysis techniques. \mbox{WASP-12} was observed in stare mode using WFC3 G141 slitless grism which contained the spectrum of both the target planetary host star and the M-dwarf binary companions to \mbox{WASP-12}, which overlapped in the spectral response on the detector. The most recent reanalysis of this data explores the effect of systematic model correction on the absolute transit level measured for the atmospheric transmission spectrum finding that the absolute transit depth is sensitive to the systematic model applied to the data (\citealt{Stevenson2014a}). 

Spectroscopically resolving transmission absorption features in transit lightcurves is important for determining compositional information of exoplanet atmospheres. Unlike \textit{Spitzer}'s IRAC instrument, which is able to obtain photometric points in H$_2$O absorption bands, HST WFC3 is vitally able to spectroscopically resolve features in the atmospheres of exoplanets. Accurately determining the parameters of a transiting planet via time series observations is dependent on robustly accounting for any systematic effects from the telescope and detector for both ground and space based instruments. Each of the models used in the literature attempt to correct the array of systematics observed independently in each dataset. However, not all datasets display the same combination of systematics, which appear to be highly dependent on the observational mode and set-up. 

In this paper we apply the marginalisation method proposed by \citet{gibson2014} to not only incorporate the analysis of multiple systematic treatments on the lightcurve but also to be applicable to multiple datasets allowing for a cross-comparison between different transmission spectra. In \S\,2 we outline the current methods used in the literature to reduce and analyse current WFC3 transit datasets. In \S\,3 we put forward and discuss the marginalisation analysis technique, which incorporates a number of previously published methods to produce robust transit parameters. We the discuss marginalisation in the context of our observations in \S\,4 and highlight the key aspects of marginalisation for transit datasets. We then contrast this technique in \S\,5, to assess the impact of different analysis methods and on the computed transit parameters.

\begin{table*}
\centering
\caption[\quad HST WFC3 G141 transmission spectral observations]{Table of exoplanets observed using WFC3 G141 grism, mode the data is observed in, the systematic model used to de-trend the transmission spectral data, author, HST program, and year of observation.}
\begin{tabular}{cccccccc}
\hline
\hline
HST program & PI & Cycle & Planet & Year & Mode & Model & Paper \\
\hline
GO\,11740 & F. Pont & Cycle 17 & HD\,189733  & 2010 & stare  & $\mathcal{GP}$ & \citet{gibson2012b} \\
GO\,12181 & D. Deming & Cycle 18 & WASP-17     & 2011 & stare  &  $\theta + doot + \delta_{\lambda}$ & \citet{mandell2013}  \\
~ 		  & ~ & ~ & HD\,209458  & 2012    & scan   & $\theta + \delta_{\lambda}(\lambda).A(\lambda)$ & \citet{deming2013}   \\
~ 		  & ~ & ~ & XO-1        & 2011    & scan   & $\theta + \delta_{\lambda}(\lambda).A(\lambda)$ & \citet{deming2013}   \\
~ 		  & ~ & ~ & HAT-P-12    & 2011   & stare  & $\theta + e^{\psi/\tau} + \phi$ & \citet{line2013}     \\
~ 	      & ~ & ~ & WASP-19     & 2011    & stare  & $\theta + doot$ & \citet{huitson2013}  \\
~ 		  & ~ & ~ & ~ & ~ & ~ & $\theta + doot + \delta_{\lambda}$ & \citet{mandell2013} \\
~         & ~ & ~ & WASP-4 & 2010 & stare & $doot$ & \citet{ranjan2014} \\
~         & ~ & ~ & TrES-2 & 2010 & stare & $doot$ & \citet{ranjan2014} \\
~         & ~ & ~ & TrES-4 & 2010 & stare & $\theta + doot$ & \citet{ranjan2014} \\
~         & ~ & ~ & CoRoT-1 & 2012 & stare & $\theta + doot$ & \citet{ranjan2014} \\
GO\,12230 & M. Swain & Cycle 18 & WASP-12 & 2011 & stare & $\theta$ & \citet{swain2013} \\
~ & ~ & ~ & ~ & ~ & ~ & $\theta + doot + \delta_{\lambda}$ & \citet{mandell2013} \\
~ & ~ & ~ & ~ & ~ & ~ & $\theta + \phi^N$ & \citet{sing2013} \\
~ & ~ & ~ & ~ & ~ & ~ & $\theta + e^{\psi/\tau} + \phi$ & \citet{stevenson2014b} \\
GO\,12251 & Z. Berta & Cycle 18 & GJ\,1214    & 2010 & stare  & $doot$ & \citet{berta2012} \\
GO\,12449 & D. Deming & Cycle 19 & HAT-P-11    & 2012 & scan   & $\theta + \delta_{\lambda}(\lambda).A(\lambda)$    & \citet{fraine2014} \\
GO\,12473 & D. Sing & Cycle 19 & WASP-31     & 2012    & scan   & $\theta + \phi^N + \delta_{\lambda}^{N}$ & \citet{Sing2015}     \\
~ 		  & ~ & ~ & HAT-P-1     & 2012    & scan   & $\theta + \phi^N$ & \citet{wakeford2013} \\
GO\,12881 & P. McCullough & Cycle 20 & HD\,189733  & 2013 & scan & $\theta$ & \citet{mccullough2014} \\
GO\,13021 & J. Bean & Cycle 20 & GJ\,1214    & 2012-13 & scan  & $\theta + e^{\psi/\tau} + \phi$ & \citet{kreidberg2014a}  \\
GO\,13064 & D. Ehrenreich & Cycle 19 & GJ\,3470 & 2013 & stare & doot & \citet{ehrenreich2014} \\
GO\,13338 & K. Stevenson & Cycle 21 & GJ\,436     & 2013 & scan & $\theta^2 + e^\phi + \phi$ & \citet{stevenson2014c} \\
GO\,13467 & J. Bean & Cycle 21 & WASP-43     & 2013    & scan   & $\theta + e^{\psi/\tau} + \phi$ & \citet{kreidberg2014b}     \\
GO\,13501 & H. Knutson & Cycle 20 & HD\,97658   & 2014    & scan   & $\theta + e^{\psi/\tau}$ & \citet{knutson2014}   \\
\hline
\multicolumn{8}{l}{$\mathcal{GP}$ - Gaussian Process; $doot$ - Divide-oot; $\theta$ - Visit-long slope; $\phi$ - HST phase; $\delta_{\lambda}$ - wavelength shift; $e^{\psi/\tau}$ - model ramp}
\end{tabular}
\label{table:WFC3}
\end{table*}

\section{Systematic model corrections}
A transit lightcurve consists of $N$ stellar flux measurements observed at time $t$, collectively referred to as the data $D$ or the lightcurve. To model each of the lightcurves we calculate a \citet{mandelagol2002} transit model $T(t,\lambda)$ following a non-linear limb-darkening law as defined in \citet{claret2000}. The model is then fit to the data by allowing the baseline flux $F_*$, R$_p$/R$_*$, and centre of transit time to vary. We fixing the other planetary system parameters such as inclination and a/R$_*$ from previously published values as the phase coverage of HST lightcurves do not well constrain these parameters on their own. The final form of the model fit to the data becomes,
\begin{equation}
F(\lambda, t) = F_* \times T(t,\lambda) \times S(t,\lambda)
\end{equation}
where $S(t,\lambda)$ is the sytematics model normalised to unity.

One of the issues encountered when analysing observational datasets is determining the impact that instrumental systematics have on the resultant measurements. Since the advent of WFC3's application to transiting exoplanets, a number of systematic models have been used to reduce G141 spectroscopic data (e.g. \citealt{berta2012}; \citealt{gibson2012b}; \citealt{wakeford2013}; \citealt{line2013}; \citealt{deming2013}; \citealt{Stevenson2014a}). 

Figure \ref{fig:WFC3_systematics} shows three examples of systematic trends observed in WFC3 transit lightcurves: ``HST breathing'' effects, visit-long slopes, and the `ramp' effect. The``HST breathing'' effect shown in Fig. \ref{fig:WFC3_systematics}a displays a periodic systematic across each orbit of data. This is attributed to the known thermal variations which occur during the orbit of HST as it passes into and out of the Earth’s shadow, causing expansion and contraction of HST. This can be most easily seen in the middle panel in relation to the HST orbital phase. The ``HST breathing'' effect systematic has been noted and corrected for in multiple datasets with a variety of parametrised models (e.g. \citealt{wakeford2013}; \citealt{line2013}; \citealt{Stevenson2014a}) which remove systematics based on functions of the HST orbital time period and phase.

Many groups have also reported a visit-long trend in WFC3 lightcurves. This trend can be seen clearly in the raw band-integrated lightcurve of HD\,209458 shown in Fig. \ref{fig:WFC3_systematics}b, which displays a significant slope across the entire observation period. This can be seen in relation to both planetary phase and exposure number. This systematic trend has not been correlated with any other physical parameter of WFC3 observations. However, it has been shown to significantly affect the resultant system parameters obtained from the lightcurve.

In addition to orbital phase trends, both in planetary and HST space, a number of lightcurves have been dominated by a systematic increase in the intensity during each group of exposures obtained between buffer dumps referred to as the `ramp' or `hook' effect (e.g. \citealt{berta2012}; \citealt{mandell2013}). The dataset of WASP-17 in Fig. \ref{fig:WFC3_systematics}c clearly displays this effect, and the bottom residual plot in terms of exposure number shows the highly repeatable aspect of the systematic. This is thought to be caused by charge trapping on the detector and it has been found that the `ramp' is, on average, zero when the count rate is less than about 30,000 electrons per pixel (\citealt{wilkins2014}). This is commonly seen in stare mode observations where the count rapidly builds up over a small pixel range. 

Strong wavelength dependent shift of the stellar spectrum across the detector throughout the course of the observation can also affect the spectroscopic lightcurves and therefore the measured transit parameters (\citealt{deming2013}). Large shifts in the wavelength direction on the WFC3 detector may introduce sub-pixel to pixel sized variations in each spectroscopic bin when dividing the spectrum into wavelength channels. This is likely the result of fast scanning rates used for very bright targets where additional motion during a rapid scan rate can introduce variations between spectral exposures, as fast scan rates spread the stellar spectrum across a larger range of the detector making it much harder for the fine guidance system to hold a fixed wavelength position on the detector. 

A combination of observational and phase dependent instrument systematics has been observed in all WFC3 transit datasets. We outline the main systematic models used to correct for these effects, with the full list of published systematic correction models presently used outlined in Table \ref{table:WFC3}. 

\subsection{Exponential model ramp}
The first method put forward to correct WFC3 systematics, as an analytical model whose parameters attempt to represent the physical processes of the instrument, was outlined by \citet{berta2012} defined as the exponential model-ramp. The exponential models apply an exponential ramp over sets of exposures, corrects for orbit-long and visit-long slopes, and is intended to model the charge trapping. \citet{line2013} show the exponential model-ramp in the form of the equation 
\begin{equation}
\frac{F_{orb}}{F_{cor}} = (C + V \theta + B \phi)(1 - R e^{\psi/\tau}) \mathrm{,}
\end{equation}
where $F_{orb}/F_{cor}$ are the lightcurve residuals, $\theta$ is the planetary phase, $\phi$ is the HST orbital phase, and $\psi$ is the phase over which the ramp feature occurs, which accounts for the visit-long slope $V$, the orbit-long slope $B$, and a vertical offset $C$ applied to the whole lightcurve. The exponential model for the ramp has an additional two parameters: the ramp amplitude $R$, and the ramp timescale $\tau$. This method is used by a number of groups when a ramp is observed in the raw band-integrated lightcurve (\citealt{kreidberg2014a}; \citealt{kreidberg2014b}; \citealt{knutson2014}).

In cases where the orbital timescale matches the phase over which the ramp occurs a simplification of the ramp-model can be used, as seen in \citet{stevenson2014c},
\begin{equation}
S(t,\lambda) =  [1 + r_0\theta + r_1\theta^2] \times [1 - e^{r_2 \phi + r_3} + r_4 \phi] \mathrm{,}
\end{equation}
where $r_{0-4}$ are free parameters and the phase $\psi$ over which the ramp feature occurs is now equal to the HST orbital phase $\phi$. This is displayed as ``$\theta + e^\phi + \phi$'' in Table \ref{table:WFC3}. \citet{fraine2014} find that a simple linear visit-long correction and a single exponential ramp in HST phase can correct for the systematics observed in the transit lightcurve of the hot Neptune \mbox{HAT-P-11} without the need for a squared term in time. 

Both of these methods rely on the timescales of each ramp to be the same in each orbit and for each orbit to have the same repeating systematic. This therefore depends heavily on the scheduling of each exposure within a HST orbit, which a number of HST WFC3 datasets do not meet.

\begin{table*}
\footnotesize{
\begin{center}
\caption[\quad Systematic model grid]{Table of all parametrised systematic models applied to the lightcurves showing the combination of visit-long trends as a function of planetary phase ($\theta$), functions of HST orbital phase ($\phi$), and functions dependent on wavelength shifts ($\delta_{\lambda}$) in the data. In addition to these models we apply the two exponential orbital phase models outlined in \citet{stevenson2014c} \label{table:grid}.}
\begin{tabular}{ccccccccccccccccccccc}
\hline
\hline
No. & $\theta$ & $\phi$ & $\phi^2$ & $\phi^3$ & $\phi^4$ & $\delta_{\lambda}$ & $\delta_{\lambda}^2$ & $\delta_{\lambda}^3$ & $\delta_{\lambda}^4$ & ~ & No. & $\theta$ & $\phi$ & $\phi^2$ & $\phi^3$ & $\phi^4$ & $\delta_{\lambda}$ & $\delta_{\lambda}^2$ & $\delta_{\lambda}^3$ & $\delta_{\lambda}^4$ \\
 \hline
0 & ~ & ~ & ~ & ~ & ~ & ~ & ~ & ~ & ~ & ~ & 25 & $\surd$ \\
1 & ~ & ~ & ~ & ~ & ~ & $\surd$ & ~ & ~ & ~ & ~ & 26 & $\surd$ & ~ & ~ & ~ & ~ & $\surd$ \\
2 & ~ & ~ & ~ & ~ & ~ & $\surd$ & $\surd$ & ~ & ~ & ~ & 27 & $\surd$ & ~ & ~ & ~ & ~ & $\surd$ & $\surd$ \\
3 & ~ & ~ & ~ & ~ & ~ & $\surd$ & $\surd$ & $\surd$ & ~ & ~ & 28 & $\surd$ & ~ & ~ & ~ & ~ & $\surd$ & $\surd$ & $\surd$  \\
4 & ~ & ~ & ~ & ~ & ~ & $\surd$ & $\surd$ & $\surd$ & $\surd$ & ~ & 29 & $\surd$ & ~ & ~ & ~ & ~ & $\surd$ & $\surd$ & $\surd$ & $\surd$ \\
5 & ~ & $\surd$ & ~ & ~ & ~ & ~ & ~ & ~ & ~ & ~ & 30 & $\surd$ & $\surd$  \\
6 & ~ & $\surd$ & ~ & ~ & ~ & $\surd$ & ~ & ~ & ~ & ~ & 31 & $\surd$ & $\surd$ & ~ & ~ & ~ & $\surd$ \\
7 & ~ & $\surd$ & ~ & ~ & ~ & $\surd$ & $\surd$ & ~ & ~ & ~ & 32 & $\surd$ & $\surd$ & ~ & ~ & ~ & $\surd$ & $\surd$ \\
8 & ~ & $\surd$ & ~ & ~ & ~ & $\surd$ & $\surd$ & $\surd$ & ~ & ~ & 33 & $\surd$ & $\surd$ & ~ & ~ & ~ & $\surd$ & $\surd$ & $\surd$ \\
9 & ~ & $\surd$ & ~ & ~ & ~ & $\surd$ & $\surd$ & $\surd$ & $\surd$ & ~ & 34 & $\surd$ & $\surd$ & ~ & ~ & ~ & $\surd$ & $\surd$ & $\surd$ & $\surd$ \\
10 & ~ & $\surd$ & $\surd$ & ~ & ~ & ~ & ~ & ~ & ~ & ~ & 35 & $\surd$ & $\surd$ & $\surd$ \\
11 & ~ & $\surd$ & $\surd$ & ~ & ~ & $\surd$ & ~ & ~ & ~ & ~ & 36 & $\surd$ & $\surd$ & $\surd$ & ~ & ~ & $\surd$ \\
12 & ~ & $\surd$ & $\surd$ & ~ & ~ & $\surd$ & $\surd$ & ~ & ~ & ~ & 37 & $\surd$ & $\surd$ & $\surd$ & ~ & ~ & $\surd$ & $\surd$ \\
13 & ~ & $\surd$ & $\surd$ & ~ & ~ & $\surd$ & $\surd$ & $\surd$ & ~ & ~ & 38 & $\surd$ & $\surd$ & $\surd$ & ~ & ~ & $\surd$ & $\surd$ & $\surd$ \\
14 & ~ & $\surd$ & $\surd$ & ~ & ~ & $\surd$ & $\surd$ & $\surd$ & $\surd$ & ~ & 39 & $\surd$ & $\surd$ & $\surd$ & ~ & ~ & $\surd$ & $\surd$ & $\surd$ & $\surd$ \\
15 & ~ & $\surd$ & $\surd$ & $\surd$ & ~ & ~ & ~ & ~ & ~ & ~ & 40 & $\surd$ & $\surd$ & $\surd$ & $\surd$ \\
16 & ~ & $\surd$ & $\surd$ & $\surd$ & ~ & $\surd$ & ~ & ~ & ~ & ~ & 41 & $\surd$ & $\surd$ & $\surd$ & $\surd$ & ~ & $\surd$ \\
17 & ~ & $\surd$ & $\surd$ & $\surd$ & ~ & $\surd$ & $\surd$ & ~ & ~ & ~ & 42 & $\surd$ & $\surd$ & $\surd$ & $\surd$ & ~ & $\surd$ & $\surd$ \\
18 & ~ & $\surd$ & $\surd$ & $\surd$ & ~ & $\surd$ & $\surd$ & $\surd$ & ~ & ~ & 43 & $\surd$ & $\surd$ & $\surd$ & $\surd$ & ~ & $\surd$ & $\surd$ & $\surd$ \\
19 & ~ & $\surd$ & $\surd$ & $\surd$ & ~ & $\surd$ & $\surd$ & $\surd$ & $\surd$ & ~ & 44 & $\surd$ & $\surd$ & $\surd$ & $\surd$ & ~ & $\surd$ & $\surd$ & $\surd$ & $\surd$ \\
20 & ~ & $\surd$ & $\surd$ & $\surd$ & $\surd$ & ~ & ~ & ~ & ~ & ~ & 45 & $\surd$ & $\surd$ & $\surd$ & $\surd$ & $\surd$ \\
21 & ~ & $\surd$ & $\surd$ & $\surd$ & $\surd$ & $\surd$ & ~ & ~ & ~ & ~ & 46 & $\surd$ & $\surd$ & $\surd$ & $\surd$ & $\surd$ & $\surd$ \\
22 & ~ & $\surd$ & $\surd$ & $\surd$ & $\surd$ & $\surd$ & $\surd$ & ~ & ~ & ~ & 47 & $\surd$ & $\surd$ & $\surd$ & $\surd$ & $\surd$ & $\surd$ & $\surd$ \\
23 & ~ & $\surd$ & $\surd$ & $\surd$ & $\surd$ & $\surd$ & $\surd$ & $\surd$ & ~ & ~ & 48 & $\surd$ & $\surd$ & $\surd$ & $\surd$ & $\surd$ & $\surd$ & $\surd$ & $\surd$ \\
24 & ~ & $\surd$ & $\surd$ & $\surd$ & $\surd$ & $\surd$ & $\surd$ & $\surd$ & $\surd$ & ~ & 49 & $\surd$ & $\surd$ & $\surd$ & $\surd$ & $\surd$ & $\surd$ & $\surd$ & $\surd$ & $\surd$ \\
  \hline
\multicolumn{21}{l}{~50 ~ $\theta$ $\times$(1 - $e^{\phi}$) + $\phi$ ~~~~~~~~~~~~~~~~~~~~~~~~~~~~~~~~~~~~~~~~~  51 ~ ($\theta$ + $\theta^2$) $\times$ (1 - $e^{\phi}$) + $\phi$} 
\end{tabular}
\end{center}
}
\end{table*}

\subsection{Polynomial models} \label{sec:polynomial}
An additional polynomial method to correct for ``HST breathing'' effects in the data is discussed in \citet{wakeford2013}, which assumes that the effects fit a polynomial function rather than being exponential in nature. This method also seeks to remove the visit-long slope observed in each WFC3 transit dataset using a linear time trend in planetary phase in addition to the HST phase corrections,
 
\begin{equation}
S(t,\lambda) = T_1\,\theta + \sum_{i=1}^{n}p_i\phi^i 
\end{equation}
where $\theta$ is planetary phase representing a linear slope over the whole visit when multiplied by the free or fixed parameter $T_1$, $\phi$ is HST phase accounting for ``HST breathing'' effects when multiplied by $p_{1-n}$, which are either free parameters or fixed to zero to fit the model to the data.

In addition to ``HST breathing'' trends, functions in wavelength shift on the detector were needed to correct for systematics seen in the lightcurve of WASP-12 (\citealt{sing2013}). The systematic model then takes the form,

\begin{equation}
S(t,\lambda) = T_1\,\theta + \sum_{i=1}^{n}p_i\phi^i + \sum_{j=1}^{n}l_j \delta_{\lambda}^j 
\end{equation}
where $\delta_{\lambda}$ is the array of the shift in the wavelength (x) direction on the detector over the visit, and $l_{1-n}$ are fixed to zero or free parameters, similar to that used for the ``HST breathing'' correction. $\delta_{\lambda}$ is created by cross-correlating the stellar spectrum for each exposure with a template stellar spectrum across the whole wavelength range. This is shown in Table \ref{table:WFC3} by  a combination of the different systematics corrected for e.g. $\theta + \phi^N + \delta_{\lambda}^{N}$. 

Importantly, this model does not require each of the orbits to have the same number of exposures, or consistent repeating systematics in each orbit or between each buffer-dump making it a robust method to apply to any transit dataset from WFC3. However, due to the large number of potential free parameters in each systematic model, and the potential to go to high orders of polynomial for each systematic, using this method can potentially introduce additional systematics if the correct model is not initially chosen.

\subsection{Divide out-of-transit}
\citet{berta2012} also suggested a separate method, called divide-oot, for correcting the systematic `ramp' or `hook' observed in a number of datasets. The divide out-of-transit method (divide-oot) relies on the hook systematic being ``extremely'' repeatable between orbits, HST phase, in a visit. 

Divide-oot uses the out-of-transit orbits to compute a weighted average of the flux evaluated at each exposure within an orbit and divides the in-transit orbits by the template created. This requires each of the in-transit exposures to be equally spaced in time with the out-of-transit exposures being used to correct them, such that each corresponding image has the same HST phase and that additional systematic effects are not introduced. While this does not rely on knowing the relationship between measured photometry and the physical state of the camera it does require there to be an even number of exposures equally spaced from orbit to orbit where the exposures occupy the same HST phase space. This is so that the systematics induced by the known ``HST breathing'' trend caused by temperature variations in its orbit can be effectively eliminated. The divide-oot method relies on the cancellation of common-mode, wavelength independent, systematic errors by operating only on the data themselves using simple linear procedures, relying on trends to be similar in the time domain over a number of orbits. This is listed as `\textit{doot}' in Table \ref{table:WFC3}.

\subsection{Spectral template corrections}
A somewhat similar technique was adopted by \citet{deming2013} and \citet{mandell2013} for their analysis of WFC3 data relying on common trends in the wavelength domain. To account for this sub-pixel change in the spectral wavelength solution for the stellar spectrum across each exposure, \citet{deming2013} introduced a spectral template technique to the extracted spectrum of each image. The template spectrum is constructed from the average observed spectrum from exposures within one hour of 1st and 4th contact. The template is then used to fit the wavelength solution to each exposure spectrum by stretching and shifting it in wavelength, stepping through small increments to determine the best fit solution. The individual spectra are then divided by the template spectrum to create residuals effectively removing any additional background contributions and the wavelength shift on the detector. This technique (`stretch and shift') also allows for the cancellation of common-mode systematics, similar to the divide-oot method, however, this requires the common-mode systematics to be consistent in wavelength rather than in time. This technique is labeled as `$\delta_{\lambda}(\lambda)$' in Table \ref{table:WFC3}. However, this method only produces a relative depth measurement for the atmospheric signatures and does not provide absolute planet-to-star radius ratio values needed for comparative studies and combined datasets. 

%
\section{Marginalisation}
Thus far it is not clear which of the systematic models that have been employed to correct WFC3 transit lightcurves is the best for each individual dataset. However, it is clear that none of these simple systematic models will work for all datasets. We attempt to rectify this by performing a marginalisation over a grid of systematic models which incorporate corrections for the different systematics observed across the exoplanet transit datasets. In this way, model selection does not have such a large influence on the final result.

\begin{figure*}
\centering 
\includegraphics[width=0.90\textwidth]{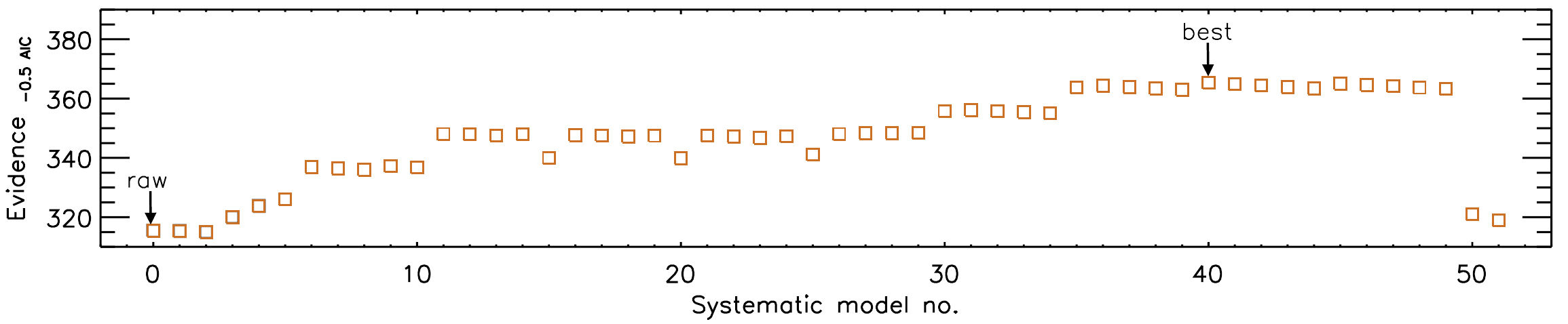}
\caption[\quad AIC evidence for each systematic model]{The evidence based on the AIC plotted against the systematic model number corresponding to Table \ref{table:grid}. This is an example of the evidence computed for WASP-31 where the best-fitting systematic model and the raw lightcurve evidence are indicated with arrows.}
\label{fig:AIC_W31_evidence}
\end{figure*}
Use of the Bayesian Information Criterion (BIC) to select between different systematic models, which takes Occam's Razor into effect by penalising models with increasing complexity, has been previously used for WFC3 observations (e.g. \citealt{wakeford2013}; \citealt{sing2013}; \citealt{Sing2015}; \citealt{haynes2015}), and has been applied to a range of datasets (e.g. \citealt{huitson2013}; \citealt{crossfield2013}; \citealt{nikolov2014}). Here we take this one step further by adopting the methodology proposed by \citet{gibson2014} of marginalising over multiple systematic models to calculate robust transit parameters. This allows us to quantify the degeneracy between our physical parameters of interest and our choice of systematic model. In this case we want to determine the value and associated uncertainty the planet-to-star radius ratio (R$_p$/R$_*$) for each of our lightcurves after correcting for the systematics inherent in the data. For each systematic model used to correct the data we calculate the evidence of fit, defined as the probability that the data would be produced given the systematic model, which is then used to apply a weight to the parameter of interest measured using that model. A weighted average of R$_p$/R$_*$ is then calculated which takes into account the individual weights of each fit and statistical likelihood of each model. This ensures that a variety of systematic models are taken into account when measuring the R$_p$/R$_*$ without having to choose between models. 

Before marginalisation, the overall systematic models that are going to contribute to the final weighting must be decided upon. We use a grid of 49 polynomial models which incorporate all known identified systematic trends in the data (see Table \ref{table:grid}), with the addition of the two exponential models outlined by \citet{stevenson2014c}, and the uncorrected lightcurve. All 52 models are then placed into an array of varying free parameters to be fit or fixed in turn and looped over for each lightcurve fit to calculate the weighting assigned to each in turn. We assume equal priors on all the systematic models tested, where no model is preferentially preferred over another.

It is also important to note that marginalisation relies on the fact that at least one of the models being marginalised over is a good representation of the systematics in the data. Our grid of parametrised models includes all combinations of factors up to the fourth order in both HST phase, to correct for ``HST breathing'' effects, and up to the fourth order in wavelength shift, in addition to the visit-long linear trend noted by all groups. By also incorporating the \citet{stevenson2014c} exponential HST phase models, with a linear and squared planetary phase trend, we make the assumption that this condition of marginalisation is satisfied. Under this approach, we effectively average the results obtained from a suite of systematic models in a principled manner. In doing so, we marginalise over our uncertainty as to which systematic model is actually the ``correct'' model. This is especially important when several models have equally well fitting systematics, as is often the case.

\subsection{Evidence and weight}\label{sec:evidence}
To calculate the weighting assigned to each of the systematic models and subsequently the final marginalised parameter for the planet-to-star radius ratio for each planetary transit, we first have to determine the evidence that each systematic model has when fit to the data. The evidence $E_q$ of fit assigned to each systematic model $S_q$ is given by the probability of the data $D$ given the model $q$ and is often referred to as the marginal likelihood. In the absence of accurate priors on which to place a likelihood, we can use an approximation for the evidence (\citealt{gibson2014}), such as,

\begin{equation}
	\ln E_q = \ln \mathcal{P}(D|S_q) \approx -\frac{1}{2} \mathrm{BIC} = \ln[\mathcal{P}(D|\alpha_{*},S_q)] - \frac{1}{2}M\ln N \mathrm{,}
\end{equation}
where the BIC is the Bayesian Information Criterion, which is equated to the logarithmic probability of the data given the parameter and systematic model ($\ln[\mathcal{P}(D|\alpha_{*},S_q)]$) minus the number of free parameters $M$ multiplied by the log number of data points being fit $N$. 

The BIC is the most commonly used criterion to select between models in the current exoplanet literature. Alternatively to the BIC, the Akaike Information Criterion (AIC) can be calculated, which does not penalise the model as strongly for added complexity given a large number of data points,
\begin{equation}
	\ln E_q = \ln \mathcal{P}(D|S_q) \approx -\frac{1}{2} \mathrm{AIC} = \ln[\mathcal{P}(D|\alpha_{*},S_q)] - M \mathrm{.}
\end{equation}

Both the AIC and BIC have strong theoretical foundations and can be used for model selection (\citealt{burnham2004}). As the number of data points in each dataset greatly exceeds that of the number of free parameters in our most complex model we choose to minimise the AIC to give our best-fitting model with the largest evidence. This is also favoured in \citet{gibson2014} as it allows for more flexible models into the likelihood, which typically leads to more conservative error estimates. 

The evidence calculated for each model additionally relies upon the uncertainty placed on the data ($\sigma$), which is dominated by photon noise in spectral extraction pipelines. To ensure appropriate uncertainties, we start by applying photon noise errorbars to each point and including our systematics models when running \emph{MPFIT} (\citealt{markwardt2009}) which uses L-M least-squares minimisation to fit the data. We then determine the best systematic model used based on the AIC and rescale the lightcurve uncertainties such that it has a reduced $\chi^2$ of 1. We then re-run each of the systematic models with the lightcurves prior to to marginalisation. Typically this rescales the errors by a factor of $\sim$1.1 to $\sim$1.2 times the theoretical photon noise limit of WFC3. Once we have applied this inflation to our errorbars the approximated evidence is modified to incorporate the likelihood function. 

To apply this to our dataset we expand the above likelihood function, 
\begin{equation}
	\ln[\mathcal{P}(D|\alpha_{*},S_q)] = \ln \left[ \prod_{i=1}^{N} \frac{1}{\sigma\sqrt{2\pi}} e^{-\frac{r_i^2}{\sigma^2}} \right] 
\end{equation}
where $\sigma$ is the uncertainty on the data, and $r_i$ represents the model residual for the $i$th datapoint. 

\begin{equation}
	\ln[\mathcal{P}(D|\alpha_{*},S_q)] = \sum_{i=1}^{N} \ln \left[  \frac{1}{\sigma\sqrt{2\pi}} e^{-\frac{r_i^2}{2\sigma^2}} \right] \mathrm{,}
\end{equation}

\begin{equation}
		= \sum_{i=1}^{N} \ln \left[ \frac{1}{\sigma\sqrt{2\pi}} \right] -\frac{1}{2} \left(\frac{r_i}{\sigma}\right)^2 \mathrm{,}
\end{equation}

\begin{equation}
	= -N\ln\left(\sigma(2\pi)^{\frac{1}{2}} \right) - \frac{1}{2}\chi^2 \mathrm{,\ leading\ to}
\end{equation}

\begin{equation}
	\ln[\mathcal{P}(D|\alpha_{*},S_q)] = -N\ln\sigma - \frac{N}{2}\ln 2\pi - \frac{1}{2}\chi^2\ \mathrm{.}
\end{equation}

Substituting Eq 11 into Eq 6, we arrive at,
\begin{equation}
	\ln E_q = - N\ln \sigma - \frac{1}{2}N\ln 2\pi - \frac{1}{2}\chi^2 - M
\end{equation}
This gives us the final form of the evidence function for each of our systematic models applied to the data. This now needs to be transformed into a weighting so that each of the systematic models is assigned a percentage of the overall probability and, when normalised, $\sum_q \mathcal{P}(S_q|D) = 1$. 

The individual weight (W$_q$) for each systematic model is calculated by
\begin{equation}
	W_q = \mathcal{P}(S_q|D) = E_q\,/\,\sum_{q=0}^{N_q} E_q \mathrm{.}
\end{equation}
Where $N_q$ is the number of models fit, $\alpha_m$ is the marginalised parameter, and $\alpha_q$ is the measured parameter for each model.

The weighting assigned to each model due to the evidence parameter can then be used to calculate the weighted mean of all the parameters ($\alpha$) of interest 
\begin{equation}
	\alpha_m = \sum_{q=0}^{N_q}(W_q \times \alpha_q) \mathrm{,}
\end{equation}
and the uncertainty ($\sigma_{\alpha}$) on that parameter can be determined from $\sigma_{\alpha_q}$ i.e. the uncertainty on the parameter $\alpha$ determined from the $q$th model,
\begin{equation}
	\sigma(\alpha)\/=\/\sqrt{\sum_{q=0}^{N}(W_q [(\alpha_q-\alpha_m)^2 + \sigma_{\alpha_{q}}^2])} \mathrm{.}
\end{equation}
This assumes that the AIC is a good approximation for the evidence calculated for each systematic model and uses a Gaussian approximation for the posterior distributions (\citealt{fruhwirth2006}). 

From the marginalisation over all 52 systematic models on the band-integrated lightcurve, the best-fitting model can reveal the dominant contributing systematics to each dataset. Figure \ref{fig:AIC_W31_evidence} shows the calculated evidence based on the AIC approximation for all 52 models when fit to an example dataset, where the systematic model with the highest overall weighting for the marginalisation is indicated by an arrow along with the raw lightcurve fit. This clearly shows the sample of systematic models that are favoured when correcting this dataset and emphasises the need for through systematic model analysis. It is sometimes the case that a number of models will have a strong weighting on the band-integrated lightcurve, where their weight\,\textgreater\,10\%. Often these models correct for the same combination of systematic trends assigning a different order to the polynomial used for correction to within 1 order. 

\begin{figure}
\centering 
\includegraphics[width=0.45\textwidth]{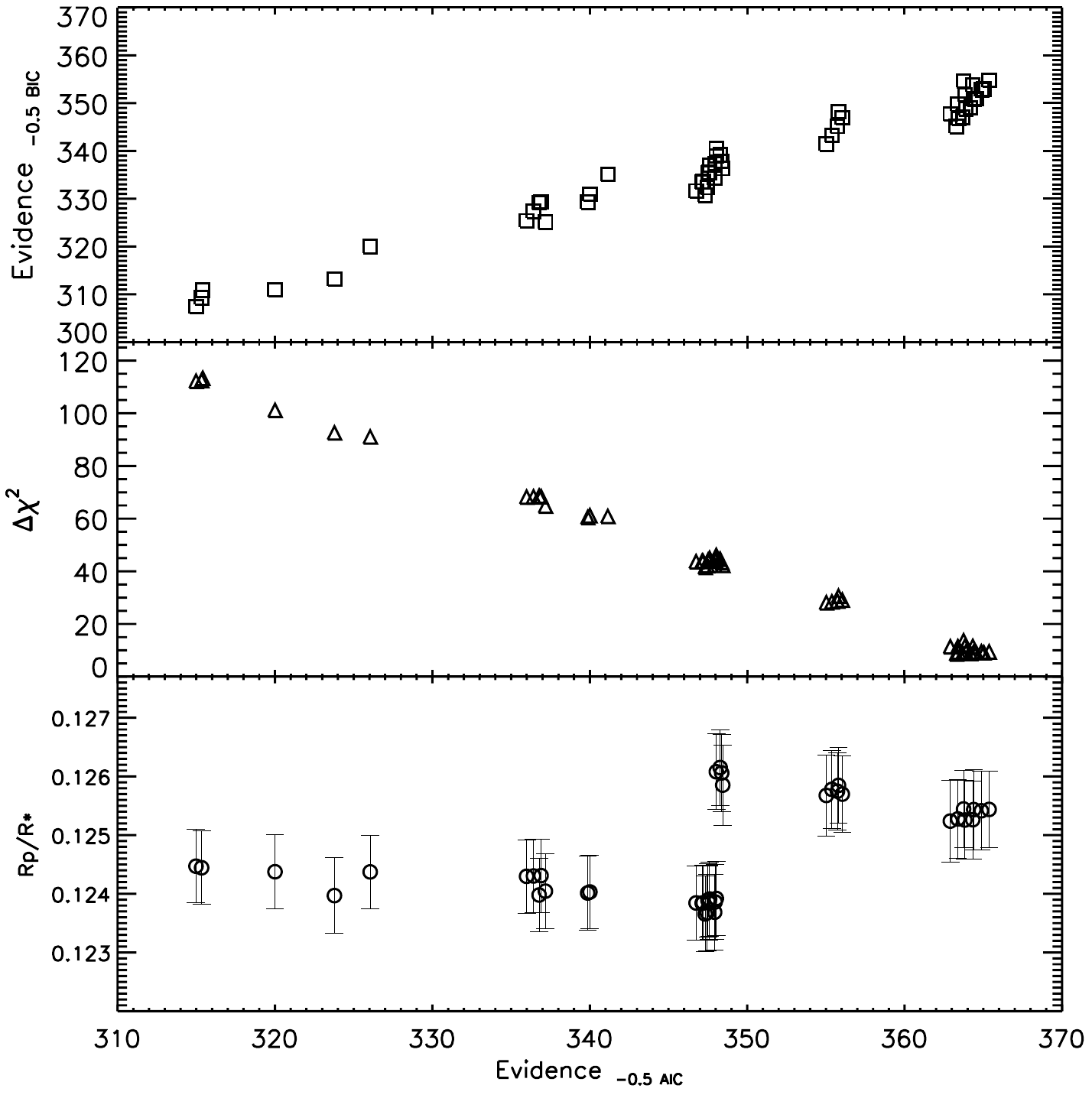}
\caption[\quad AIC evidence correlations]{The evidence based on the AIC plotted against the evidence based on the BIC (top-squares), the $\chi^2$ (middle-triangles), the R$_p$/R$_*$ (bottom-circles) for the 52 models computed for WASP-31b. The best-fitting model is that with the highest AIC evidence parameter, while a similar trend is seen by reducing the $\Delta\chi^2$ the AIC penalises the models for added complexity. }
\label{fig:AIC_evidence}
\end{figure}
In Fig.\,\ref{fig:AIC_evidence} we demonstrate the statistical correlation between different factors used to select models, comparing the evidence based on the AIC approximation to that of the evidence based on the BIC approximation, and to the $\Delta\chi^2$ for all 52 systematic models. We additionally show the difference in the measured R$_p$/R$_*$ computed for each model relative to the weight applied to that fit, again demonstrating the importance of using the correct systematic model when correcting transit data. 

%
\section{Observations}
Observations of single transit events are conducted using Hubble Space Telescope's (HST) Wide Field Camera 3 (WFC3) in the infrared (IR) with the G141 spectroscopic grism. Our observations span two HST large programs and two observation modes, stare mode and spatial scan mode, acquired between 2011--2012 over 25 HST orbits. 
\begin{table*}
\centering
\caption[\quad HST WFC3 G141 observation parameters]{Table of the observation parameters for the five planetary transits measured with WFC3. The planetary system parameters used for each of the datasets is also listed along with the band-integrated limb-darkening parameters calculated using 1D Kurucz stellar models.
}
\begin{tabular}{lccccc}
\hline
\hline
~ & HAT-P-1 & WASP-31 & XO-1 & HD\,209458 & WASP-17  \\
\hline
GO Program & 12473 & 12473 & 12181 & 12181 & 12181 \\
Date & 2012--07--05 & 2012--05--13 & 2011--09--30 & 2012--09--25 & 2011--07--08  \\
Mode & Scan & Scan & Scan & Scan & Stare \\
NSAMP & 4 & 8 & 9 & 5 & 16  \\
Subarray size & 512 & 256 & 128 & 256 & 512  \\
Exposure Time (s) & 46.695 & 134.35 & 50.382 & 22.317 & 12.795  \\
Peak Pixel Count & 25,000 & 38,000 & 38,000 & 44,000 & 64,000  \\
No. Exposures & 111 & 74 & 128 & 125 & 131 \\
Scan Rate (pix/s) & 1.07 & 0.15 & 0.43 & 7.43 & - \\
\hline
\hline
Rp (R$_J$) & 1.319 & 1.55 & 1.209 & 1.38 & 1.932 \\
Mp (M$_J$) & 0.525 & 0.48 & 0.942 & 0.714 & 0.477 \\
T$_{eff}$ (K) & 1322 & 1570 & 1210 & 1459 & 1755 \\
Period (days) & 4.46529976 & 3.405886001 & 3.94163 & 3.52474859 & 3.7354833692\\
Epoch (MJD) & 53979.43202 & 56060.69042 & 55834.3419 & 56195.7595 & 55750.2973239 \\
inclination ($^\circ$) & 85.634 & 84.670 & 88.92 & 86.59 & 86.917160 \\
a/R* & 9.91 & 8.19 & 11.24 & 8.859 & 7.03 \\
T* (K) & 5980 & 6250 & 5750 & 6117 & 6550 \\
$[Fe/H]*$ & 0.130 & -0.2 & 0.02 & -0.02 & -0.25 \\
\hline
\multicolumn{6}{c}{Best-fit band-integrated limb-darkening coefficients}\\
c1 & 0.58115522 & 0.49271309 & 0.57505270 & 0.55407671 & 0.46913939 \\
c2 & 0.08416305 & 0.32553801 & 0.07068500 & 0.15318674 & 0.38874433 \\
c3 & -0.1886652 & -0.5299505 & -0.1131334 & -0.2741835 & -0.6396845 \\
c4 & 0.05957531 & 0.21191308 & 0.01649965 & 0.09583456 & 0.26591443 \\
\hline
\end{tabular}
\label{table:observation_parameters}
\end{table*}

\begin{figure}
\centering 
\includegraphics[width=0.45\textwidth]{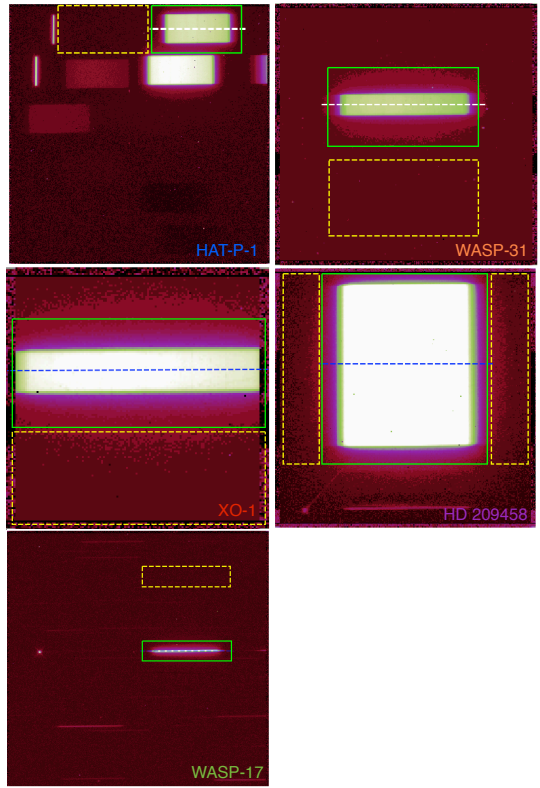}
\caption[\quad Exposure ima frame for each planet]{Single exposure ima output image for each observed exoplanet host star. Each spectral exposure is outlined in green with the center of the spectrum marked with a dotted line. The region used for background subtraction is boxed by a dashed yellow line. }
\label{fig:ima_exposures}
\end{figure}
Table \ref{table:observation_parameters} outlines the observational and planetary system parameters for each of the exoplanet hot Jupiters studied here: \mbox{HAT-P-1b}, \mbox{WASP-31b}, \mbox{XO-1b}, \mbox{HD\,209458b}, and \mbox{WASP-17b}. Each of these exoplanet transmission spectra has been previously analysed and published: \mbox{HAT-P-1b} (\citealt{wakeford2013}), \mbox{WASP-31b} (\citealt{Sing2015}), \mbox{XO-1b} (\citealt{deming2013}), \mbox{HD\,209458b} (\citealt{deming2013}), and \mbox{WASP-17b} (\citealt{mandell2013}). Across these five exoplanet transmission spectra there are a variety of measured features over the expected H$_2$O absorption bands, from full amplitude features extending several scale heights in the atmosphere, to muted and absent features. The study here is intended to demonstrate the most uniform analysis and comparison to date, such that differences between spectra can be more easily tied to the planets themselves and changes between different reduction techniques can be minimised. 

We use the ``{\it ima}" outputs from WFC3's \emph{Calwf3} pipeline. For each exposure, \emph{Calwf3} conducts the following processes: bad pixel flagging, reference pixel subtraction, zero-read subtraction, dark current subtraction, non-linearity correction, flat-field correction, and gain and photometric calibration. The resultant images are in units of electrons per second. A single exposure for each of the five targets are shown in Fig. \ref{fig:ima_exposures}, with the target spectrum outlined. This figure already demonstrates the difference in the individual observation strategies used for WFC3. The larger subarrays used for HAT-P-1 and WASP-17 include both the 0th and 1st order spectrum, while the smaller subarrays only contain the 1st order spectrum. There is also a clear difference between the brightest target, HD\,209458 (V mag = 7.7), and dimmer targets (V mag \textgreater 10.3), where a much larger scan area is needed for bright targets so that the detector is not saturated during the exposure. We also note that each HAT-P-1 exposure includes the spectral trace of the companion star to the target exoplanet host star.
\begin{figure}
\centering 
\includegraphics[width=0.45\textwidth]{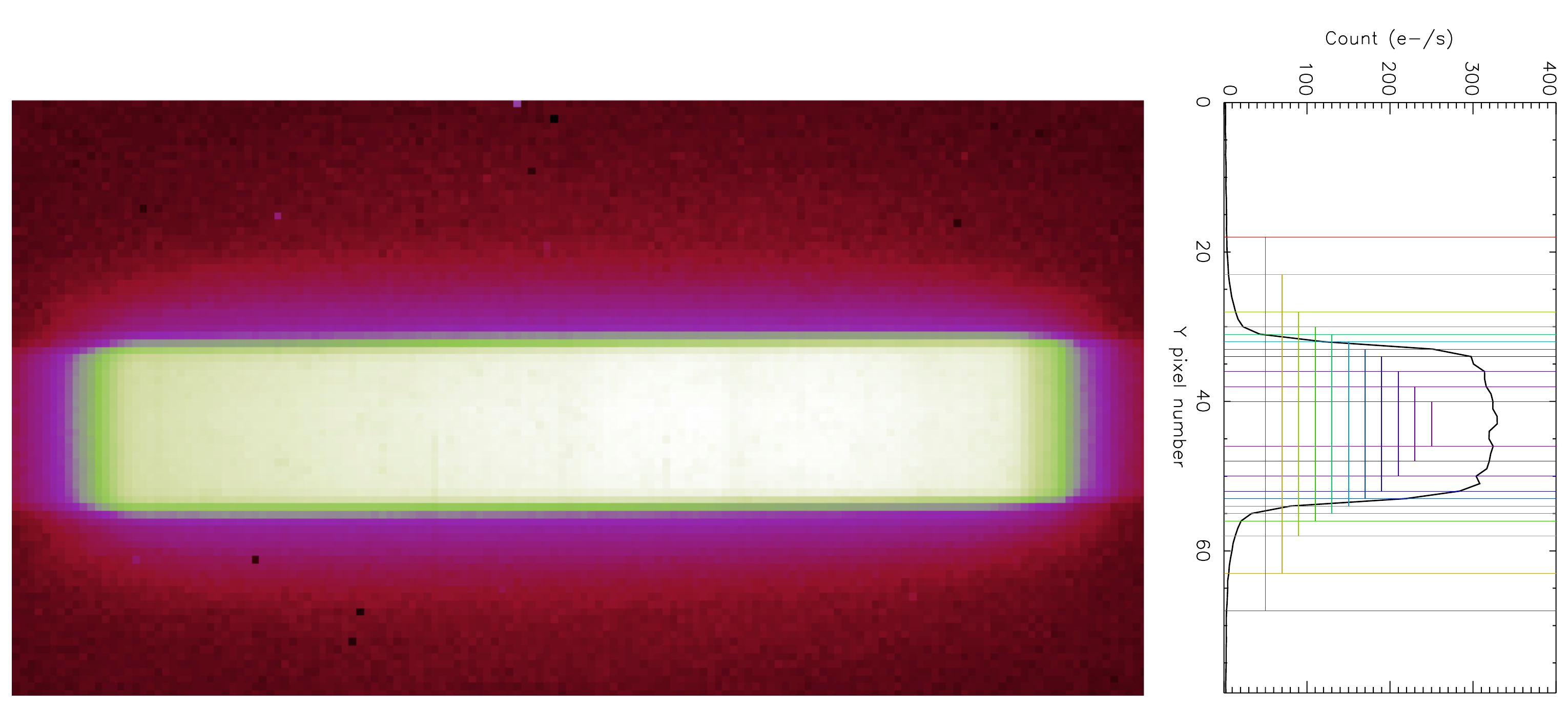}
\caption[\quad Spectral extraction and aperture determination]{Left: example exposure in spatial scan mode from the ``ima'' output. Right: The spectral trace in the cross-dispersion direction from one pixel column is shown in black, with a series of aperture sizes marked out by the colored brackets. The best aperture size is determined by minimising the standard deviation of the lightcurve residuals when fit with a standard \citet{mandelagol2002} transit model from the base planetary parameters.}
\label{fig:aperture_center}
\end{figure}

For each of our five exoplanetary transit datasets we extract each exposure spectrum with our custom \emph{IDL} routine \textit{spextract}, which optimises the aperture over which the target spectrum is exposed on each image (see Fig. \ref{fig:aperture_center}). We then compute the band-integrated lightcurve by summing the flux over all exposed wavelengths to obtain the planet-to-star radius ratio (R$_p$/R$_*$) and centre of transit time by fitting a \citet{mandelagol2002} transit model, created with non-linear limb-darkening parameters, to our data using the \emph{IDL} code \emph{MPFIT}. Due to the limited phase coverage of the HST transit observations, when fitting the band-integrated and spectroscopic lightcurves for R$_p$/R$_*$ we fix the system parameters such as inclination and a/R$_{*}$ from a joint fit between HST STIS, WFC3, and \textit{Spitzer} IRAC using a Markov Chain Monte Carlo (MCMC) analysis (\citealt{sing2016}), as further constraints cannot be placed with these individual datasets.

For each of the spectroscopic transit observations we compute the band-integrated lightcurve and transmission spectrum from 1.1--1.7\,$\mu$m. Here we present the results from each of the hot Jupiter exoplanet transit lightcurves together, discussing the data with respect to the analysis technique and individual observation strategies. 

\begin{figure*}
\centering 
\includegraphics[width=0.90\textwidth]{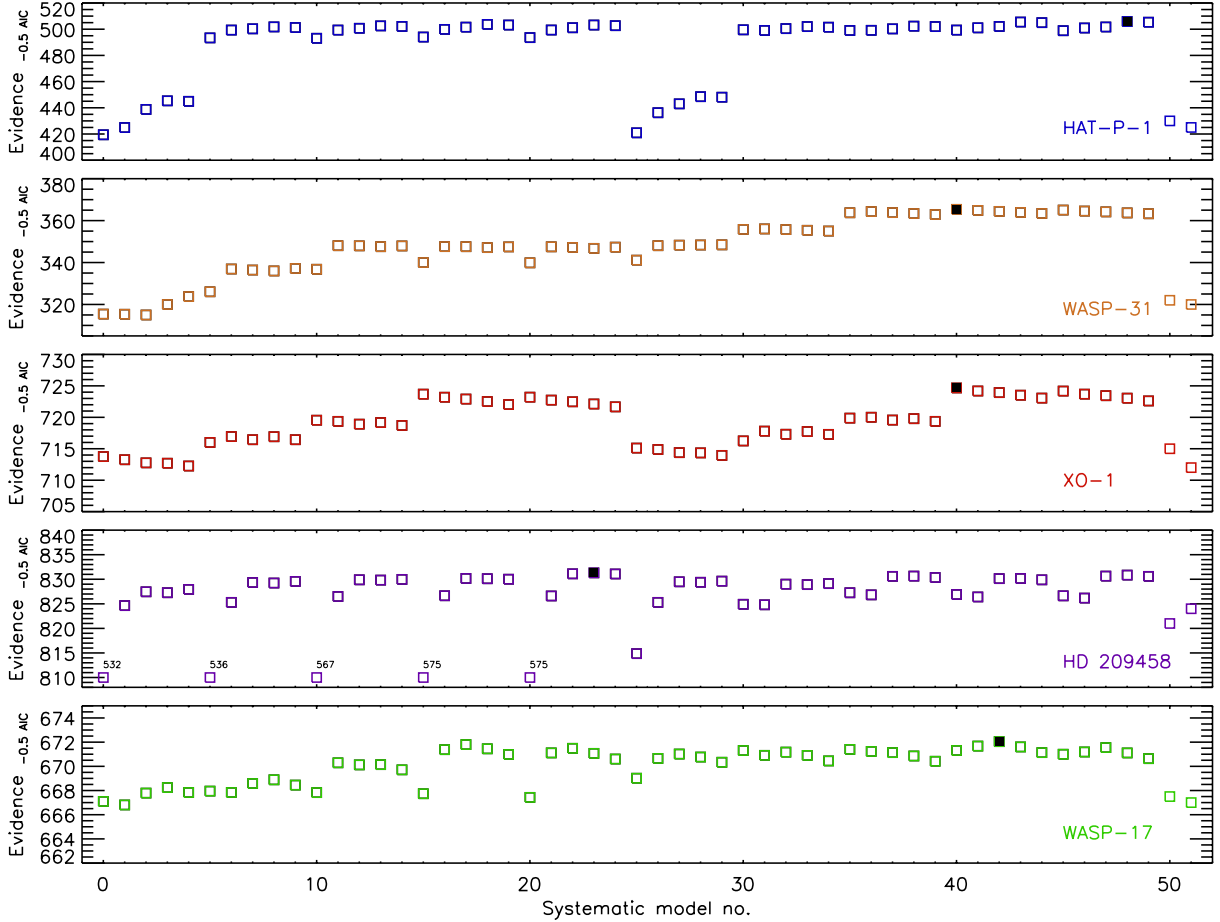}
\caption[\quad Evidence based on the AIC for each band-integrated lightcurve]{The evidence based on the AIC approximation for each systematic model applied to the band-integrated lightcurve for each exoplanet transit. Top to Bottom: \mbox{HAT-P-1} (blue), \mbox{WASP-31} (orange), \mbox{XO-1} (red), \mbox{HD\,209458} (purple), \mbox{WASP-17} (green). The best-fitting systematic model for the band-integrated lightcurve is filled in with a black square for each planet in each plot. A number of point on the HD\,209458 plot have been artificially shifted on the scale shown with their original values listed above the shifted points as they are strongly disfavoured by the data. }
\label{fig:AIC_evidence_all}
\end{figure*}
\begin{figure*}
\centering 
\includegraphics[width=0.95\textwidth]{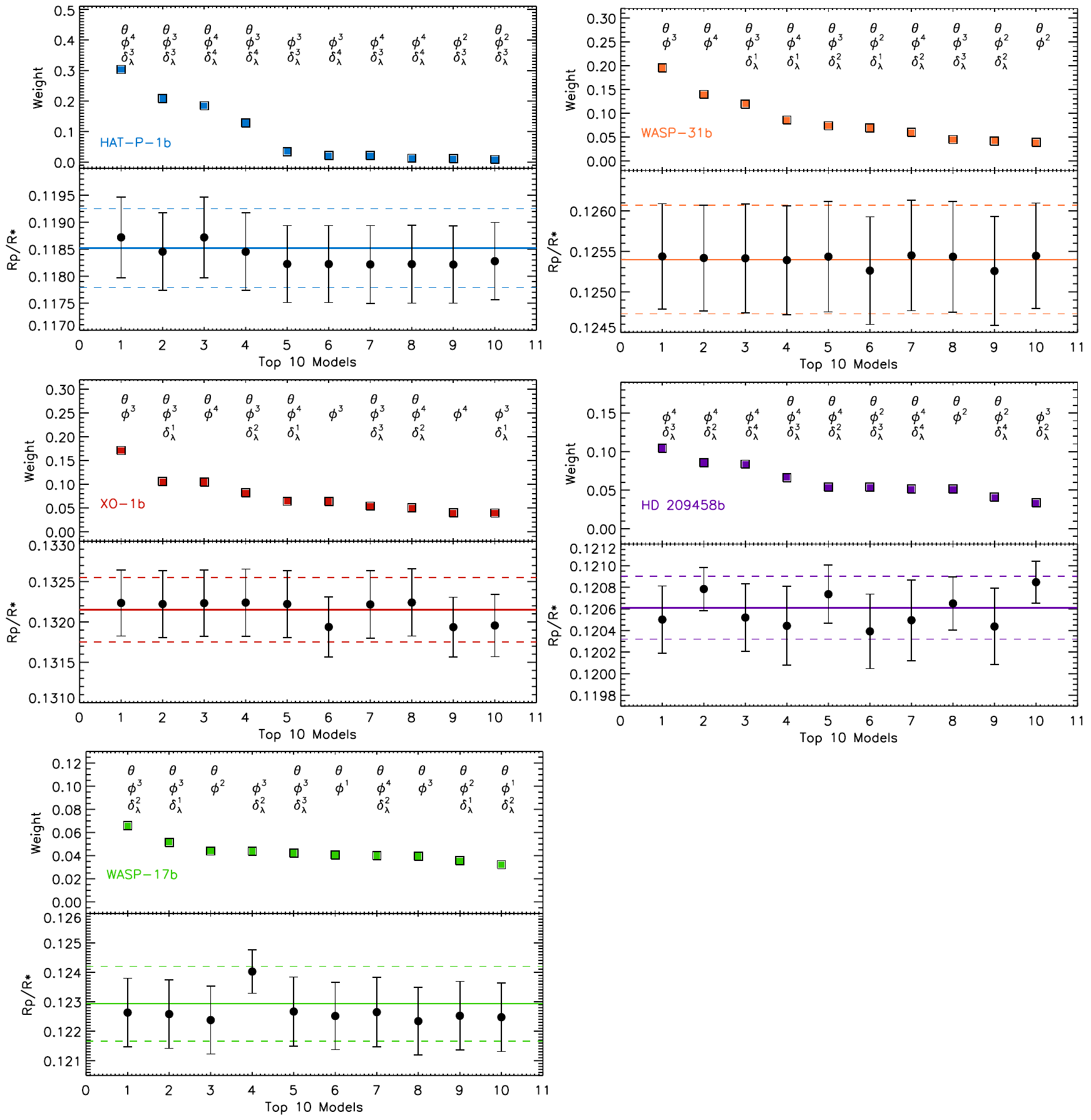}
\caption[\quad Evidence based on the AIC for the top ten systematic models with associated lightcurve parameter]{For each planetary dataset; HAT-P-1b (blue, top left), WASP-31b (orange, top right), XO-1b (red, middle left), HD\,209458b (purple, middle right), WASP-17b (green, bottom left) - Top: Weighting for the top 10 models fit to the band-integrated lightcurve based on the AIC approximation. The model parameters are outlined below each model with best to worst from left to right. $\theta$ corrects a visit-long slope, $\phi^N$ the HST orbital phase, where $N$ is the order of the polynomial used, and $\delta_{\lambda}^{N}$ is the wavelength shift polynomial correction applied. Bottom: The computed R$_p$/R$_*$ and uncertainty for each of the models as in the top plot. The solid horizontal line represents the final marginalised radius ratio with the dashed lines marking the uncertainty range.}
\label{fig:AIC_weight_all}
\end{figure*}

\subsection{Band-integrated lightcurve marginalisation}
We use our systematic grid to determine the impacting systematics for each transit dataset. The grid consists of 52 systematic models using a combination of systematics in the polynomial model and the two different exponential models (see Table \ref{table:grid}). We then perform a marginalisation on the computed planet-to-star radius ratio, and centre of transit time, and determine the best-fitting systematic model with the maximum weight from the band-integrated lightcurve. 

For each of our datasets we use marginalisation as a tool to compute robust transit depths from the band-integrated lightcurve. This also allows us to determine the main systematics impacting the lightcurves from the different spectral targets in a common and directly comparative manner. Figure \ref{fig:AIC_evidence_all} shows the evidence based on the AIC parameter for each systematic model used to correct the band-integrated lightcurve following the model numbers in Table \ref{table:grid}. We highlight the model favoured by this criterion corresponding to the systematic corrections with the highest weighting. From each of the fitting statistics it becomes clearer to see the differences across the datasets. For HAT-P-1 there is little difference between a large number of the systematic models with distinct groups of models which correct for only wavelength shifts and no ``HST breathing'' effects being disfavoured in the final marginalisation. Interestingly, the band-integrated lightcurve of WASP-31 does not favour systematic models which only correct for the known ``HST breathing'' effect, while the introduction of an additional visit-long linear correction makes these models most favourable. The systematic model correction for XO-1 favours higher order polynomials across HST phase and wavelength shift with little difference incurred with or without the visit-long slope. HD\,209458 strongly disfavours corrections with only the ``HST breathing'' effect, with the evidence based on the AIC approximation several hundred points below the other systematic correction fits (these points have been artificially shifted on the scale shown with their original values listed above the shown points). This makes the other systematic models relatively stable in that no model appears to be greatly favoured above any others. 

For WASP-17b it can clearly be seen in the raw lightcurve that there is a strong ‘hook’ feature present (Fig. \ref{fig:WFC3_systematics}), with an orbit-to-orbit repeating pattern in the residuals. The grid of systematic models used to correct the white lightcurve do not accurately account for this ‘hook’. To effectively correct for this systematic in the white lightcurve we use the divide-oot routine discussed in section 2.3, which removes common- mode time-dependent systematics. We then additionally apply a marginalisation over the systematic grid to remove any additional time- independent systematics and determine the planet-to-star radius ratio, where most models incorporating $\theta$, $\phi$, and $\delta_{\lambda}$, show some favourability. With those only correcting for ``HST breathing'' trends again least favoured in each group of models applied to the data. 

We also note that the two exponential models for each of the lightcurves are not heavily favoured and likely contribute negligibly to the resultant transit parameters. This may be the result of equal priors being placed on all the models tested. While this assumption holds when considering models from within the same family (i.e polynomial expansions), seperate priors may need to be placed when combining information from two different families of models. However, given that the exponential models were always found to give negligible weights, they are not likely sensitive to the priors.

\begin{figure*}
\centering 
\includegraphics[width=0.95\textwidth]{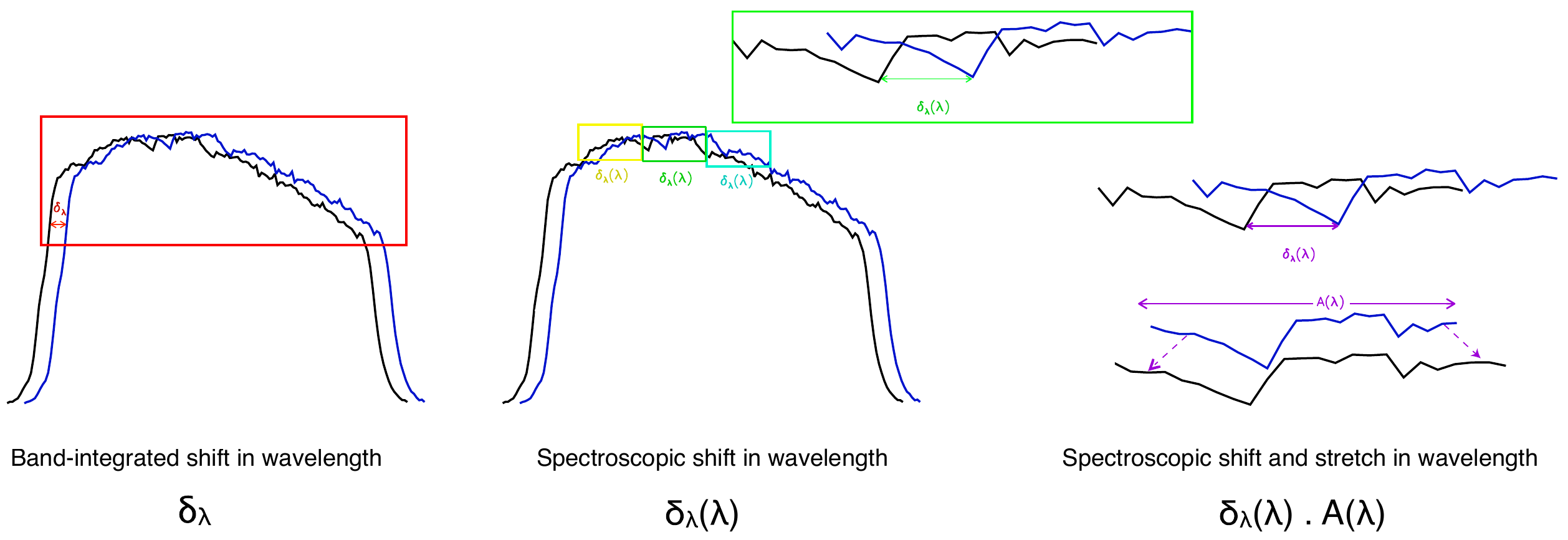}
\caption[\quad Wavelength shift diagrams]{Diagram showing the three different methods for addressing the shift in the spectral wavelength position on the detector over the course of the observations. Band-integrated shift in wavelength, $\delta_\lambda$, calculated using the whole spectral range. Spectroscopic shifts in wavelength, $\delta_\lambda(\lambda)$ calculated for each wavelength bin. Spectroscopic shift and stretch in wavelength, $\delta_\lambda(\lambda).A(\lambda)$, calculated using both a shift and stretch in the spectral template to correct systematic shifts on the detector during the observation.\\}
\label{fig:wavelength_shift_diagram}
\end{figure*}

Marginalisation offers a transparent analysis method where each stage can be deconstructed and used to understand both the dataset and the instrument. Figure \ref{fig:AIC_evidence_all} shows the top ten models favoured for each of the transit datasets analysed and the resultant planet-to-star radius ratio following the L-M least-squares fitting (L-M) with the \citet{mandelagol2002} transit model. This shows more clearly the impact the different systematic models have on the desired transit parameters. When the weighting assigned to the systematic model becomes negligible, the impact on the final values also reduces, meaning only the best models become relevant in the final parameter determination. From this it is also possible to determine the main source of corrections associated with different models. For example, the absolute radius ratio computed from the WASP-17b data indicates that the visit-long correction impacts both the calculated radius ratio and the uncertainty associated with the fit. The visit-long correction is especially important for this dataset to account for the absence of post-transit data which is used to determine the stellar baseline flux. By using the results computed from all of the systematic models that may best represent the data, marginalisation produces a robust and accurate description of the specific dataset and thus the planetary parameter, as can be seen in the final band-integrated radius ratio and the associated uncertainty, shown by the colored lines in each plot. 

Similar to our previously published work (e.g. \citealt{wakeford2013}; \citealt{nikolov2014}; \citealt{Sing2015}), to corroborate the uncertainties estimated by L-M using MPFIT, we also applied a MCMC analysis, using the IDL routine EXOFAST (eastman2012a), to the band-integrated light curves for the five most favored systematic models in each planetary dataset. While, MPFIT uses L-M least squares fitting the parameter errors from the covariance matrix calculated using numerical derivatives, the MCMC evaluates the posterior probability distribution for each parameter of the model. Using the MCMC analysis we find that the uncertainties are within 10\% of those calculated using our analysis with L-M, as they are largely gaussian in nature. We calculate the mean percentage difference between the MCMC and L-M analysis, for the uncertainties associated with the measured Rp/R*, calculated using the top five systematic models on the band-integrated lightcurve for each dataset, HAT-P-1b, WASP-31b, XO-1b, HD\,209458b, and WASP-17b respectively, 1.6\%, 7.0\%, 6.1\%, 4.8\%, and -16.4\%. In the case of WASP-17b the MCMC solution produces slightly larger uncertainties on average compared to MPFIT. This slight overestimation in the MCMC is likely the result of separately fitting the residual systematics with the model and divid-oot, whereas when using MPFIT we fit these simultaneously.

Marginalisation is therefore applicable to a range of datasets where the general systematics that may impact the data are known but not well understood for each observation. By computing the desired parameter for each systematic model and marginalising over all results according to their likelihood, information is preserved. This is important for transit lightcurves where the absolute radius ratio is important for combining data from different instruments (i.e STIS, WFC3, Spitzer), and when conducting comparative studies between different exoplanet atmospheres.

%
%
%
%
\subsection{Spectroscopic lightcurve marginalisation}
In addition to performing marginalisation on the band-integrated lightcurves we also compute the evidence and weighting for each systematic model applied to the individual spectroscopic lightcurves to test the impact on the calculated transmission spectrum. 

We compute the spectroscopic transmission spectrum by binning up each exposure spectrum into a number of wavelength bins. We perform sigma clipping on each of the datasets using the best-fitting band-integrated lightcurve systematic model from the grid shown in Table \ref{table:grid}, to remove wavelength dependent outliers that deviate from the residual scatter by greater than 4$\sigma$. 
We then fit the transit lightcurve in each bin with our systematic grid to determine any wavelength dependent systematics in each bin. We then compute the marginalised transit parameters for each spectroscopic lightcurve and thus the final transmission spectrum. 

The grid of systematic models applied to each dataset is based on three main systematics observed in WFC3 time-series data. Of these three systematics, visit-long slope, HST thermal variations, and wavelength shift, only the shift in wavelength on the detector is based on a wavelength dependent array. As outlined in section \ref{sec:polynomial} the array for the shift in wavelength over the course of the observation is calculated by cross-correlating the stellar spectrum with a template spectrum. This array can be calculated for the whole stellar spectral band, or calculated for each spectroscopic bin using a wavelength defined section of the spectral trace. In the following sections we analyse the transmission spectrum of our five targets using the band-integrated wavelength shifts ($\delta_\lambda$), and the individual spectroscopic wavelength shifts ($\delta_\lambda(\lambda)$), using marginalisation to evaluate the impact of the systematics. We then apply a spectroscopic shift and stretch ($\delta_\lambda(\lambda).A(\lambda)$) method to determine the robust nature of transmission spectral structures in each planetary dataset (see Fig. \ref{fig:wavelength_shift_diagram}) and present marginalisation as the best technique for measuring planetary transmission spectra.

\subsubsection{Band-integrated wavelength shifts}
\begin{figure}
\centering 
\includegraphics[width=0.47\textwidth]{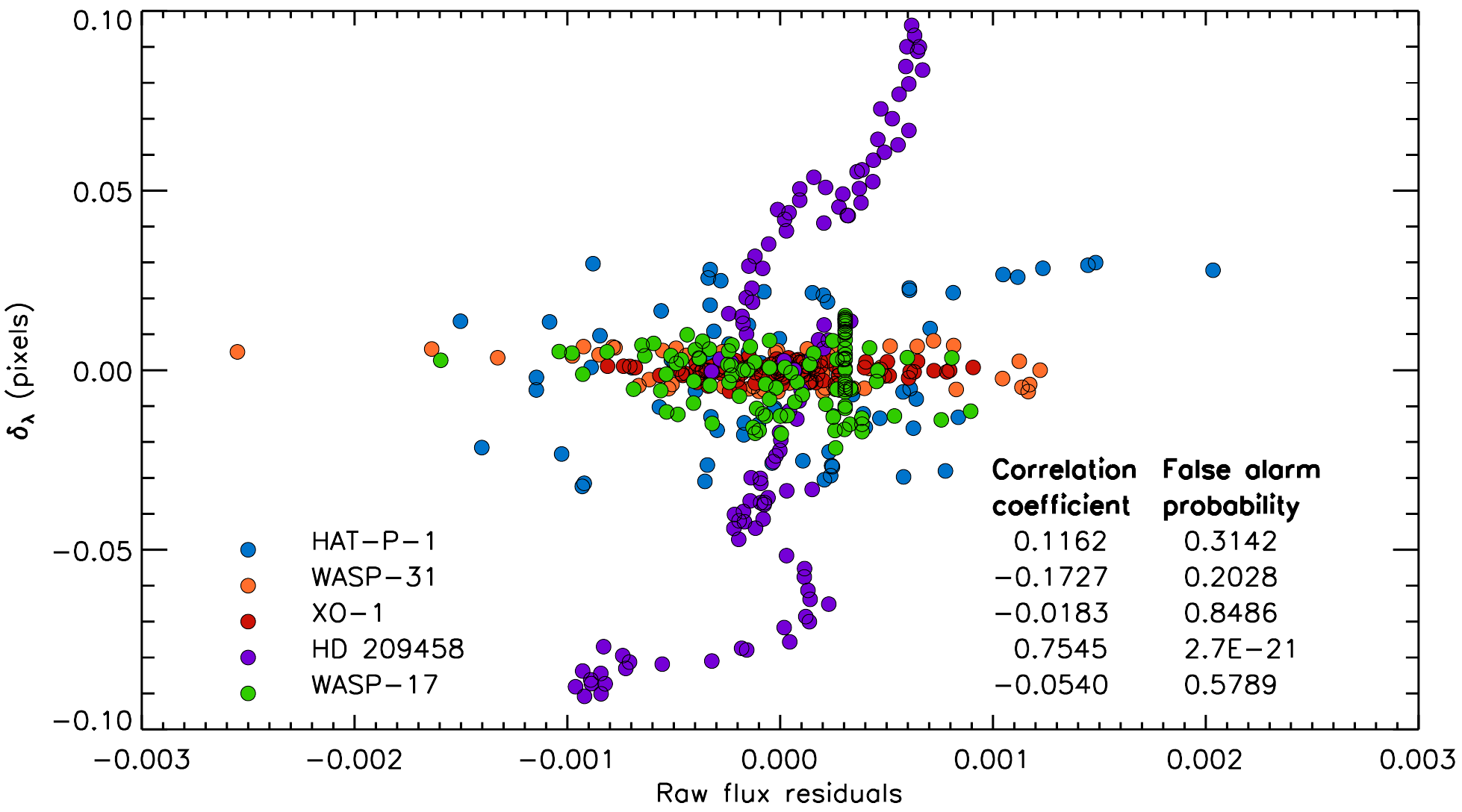}
\caption[\quad Wavelength shift vs. flux]{Trends in band-integrated wavelength shift ($\delta_\lambda$) in correlation to the observed stellar flux, where the transit is removed. The correlation coefficients show that the $\delta_\lambda$ is weakly correlated with all of our datasets apart from HD\,209458b, which shows a strong correlation (0.7545) with a negligable false alarm probability. \\
}
\label{fig:wavelength_correlation}
\end{figure}

As shown by the highest weight models from the band-integrated lightcurves (Fig. \ref{fig:AIC_weight_all}) shifts in the wavelength position of the spectrum across the course of the observation can play a significant role in the lightcurve systematics. To determine the impact of the physical shift on the detector across the whole spectral trace, we compare the wavelength shift to the raw flux residuals computed from the band-integrated lightcurve ($\delta_\lambda$), and calculate the correlation coefficient and false alarm probability for each dataset as measured by the Spearman's correlation coefficient. Figure \ref{fig:wavelength_correlation} shows that there is no generic correlation between the shift in wavelength on the detector and the systematics observed in the band-integrated lightcurve for almost all datasets considered, although some datasets do show significant trends. This highlights the need to investigate the correlation between wavelength shift on the detector and the systematics present in the band-integrated lightcurve, as the effect could be significant in order to detrend the data. 

The strong correlation between wavelength shifts and the raw data measured for \mbox{HD\,209458}, which was also noted by \citet{deming2013}, suggests that wavelength dependent systematics are dominant in these observations. This correlation is also shown by the most favoured systematic model from the band-integrated lightcurve based on the weighting calculated prior to marginalisation. The spectrum of HD\,209458b differs significantly from the other targets in this sample as it is a much brighter target, roughly 3 magnitudes. This requires a much higher scan rate over a larger range, which likely leads to the larger shifts measured. We investigate the impact this large wavelength shift has on the computed transmission spectrum and the use of different techniques related to the wavelength shift to account for this noted systematic effect.  

Similar to the analysis conducted on the band-integrated lightcurve, we calculate the R$_p$/R$_*$ and uncertainty by fitting each spectroscopic lightcurve with all 52 systematic models using the $\delta_\lambda$ array. We then margainalise over all models in each spectroscopic bin using the MLE based on the AIC to produce the atmospheric transmission spectrum for each planet. This assumes that the shift in wavelength position on the detector is constant across the whole spectral trace with no additional systematic shifts in each individual wavelength regime selected by binning the data. Again, by using marginalisation in each bin separately, we allow the data to define the systematic corrections being applied to each lightcurve and then use all of the information to calculate the final transmission spectrum. In each case this produces a marginal amount of additional scatter in the computed R$_p$/R$_*$ for each bin while maintaining a robust transmission spectral shape across the whole spectral range.

\begin{figure}
\centering 
\includegraphics[width=0.47\textwidth]{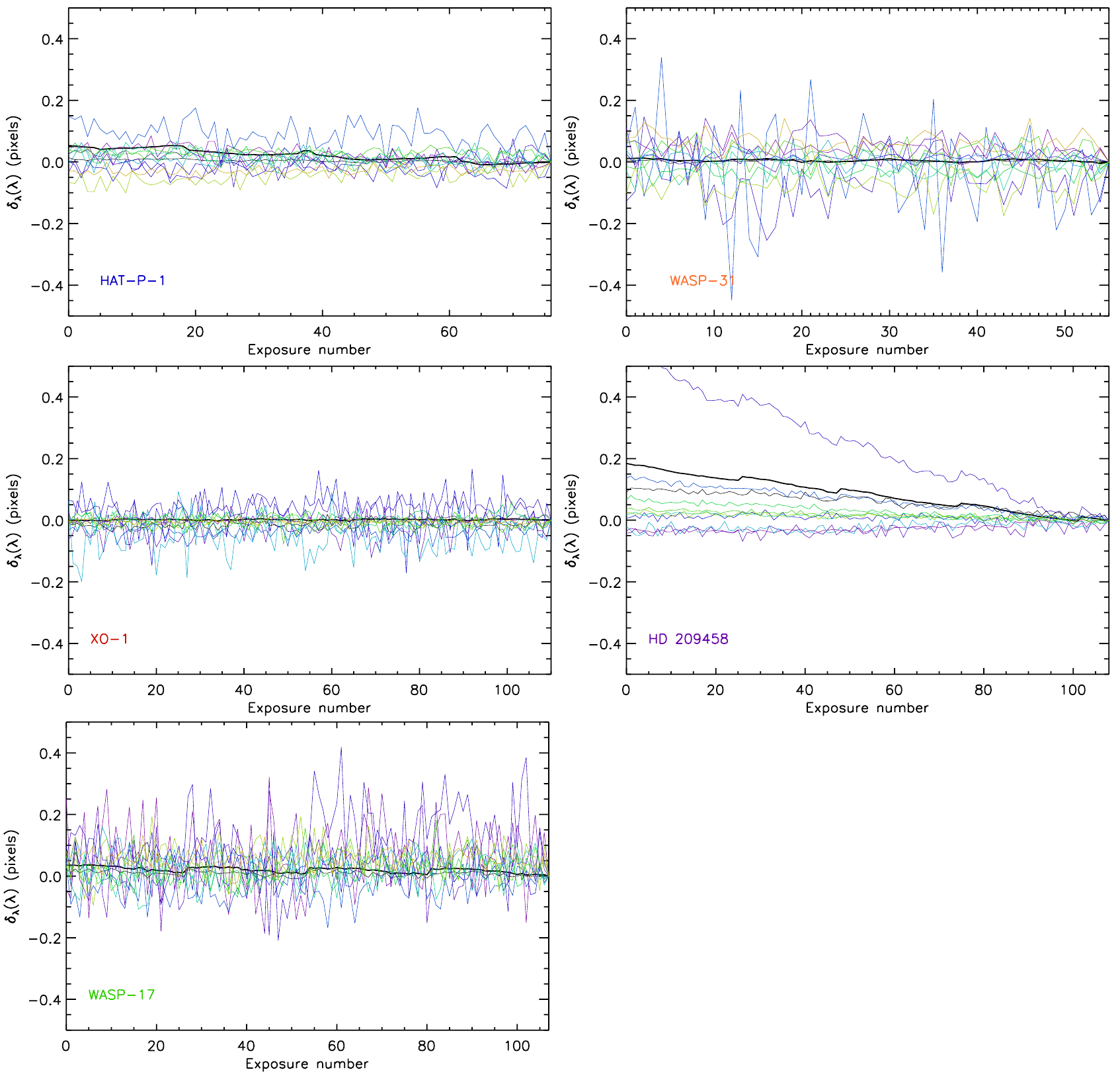}
\caption[\quad Spectroscopic wavelength shifts]{A plot of the wavelength shift measured in each spectroscopic bin of the observed stellar spectrum, $\delta_\lambda(\lambda)$, when cross-correlated with a template spectrum for each bin. The wavelength shift measured using the whole spectrum is shown as a thick black line. Each planetary dataset is represented in a seperate plot. The wavelength shifts are shown for spectroscopic bins spanning 10 pixels.}
\label{fig:wavelength_shifts}
\end{figure}

\subsubsection{Spectroscopic wavelength shifts}
To determine if there is a wavelength dependence to the shifts on the detector, we calculate an array of shifts for each spectral bin, $\delta_\lambda(\lambda)$. Each $\delta_\lambda(\lambda)$ array is calculated in the individual wavelength channels by cross-correlating each portion of the spectrum with a template of the same region of the stellar spectrum for that spectroscopic channel. A similar method is used in the analysis of WASP-33b emission spectra \citet{haynes2015}, which showed additional wavelength dependent systematics associated with the shift of the spectral trace in the wavelength position on the detector. 

Figure \ref{fig:wavelength_shifts} shows the wavelength shift calculated for each exposure, by cross-correlating the spectra for bins spanning 10 pixels compared to the wavelength shift across the whole spectrum shown as a bold black line. This shows that in a majority of the datasets, the shift in wavelength is consistent across the spectrum and across each exposure, as shown from the correlation in Fig. \ref{fig:wavelength_correlation}. However, the spectrum of HD\,209458b shows significant shifts in each wavelength bin, similar to the shifts in the full spectrum, with additional changes between the different spectral channels. Again this may be the result of the fast scan rate used in the observations of HD\,209458 as it is a much brighter star. 

To test the effect these have on the resultant transmission spectrum, we use the wavelength shift for each channel for the polynomial models in our systematic grid and applied it to each bin (i.e. replacing our $\delta_\lambda$ array with an array of $\delta_\lambda(\lambda)$ for each wavelength bin). We then marginalise over the whole of our systematic grid to produce the final transmission spectrum across the spectral range.

\begin{figure*}
\centering 
\includegraphics[width=0.95\textwidth]{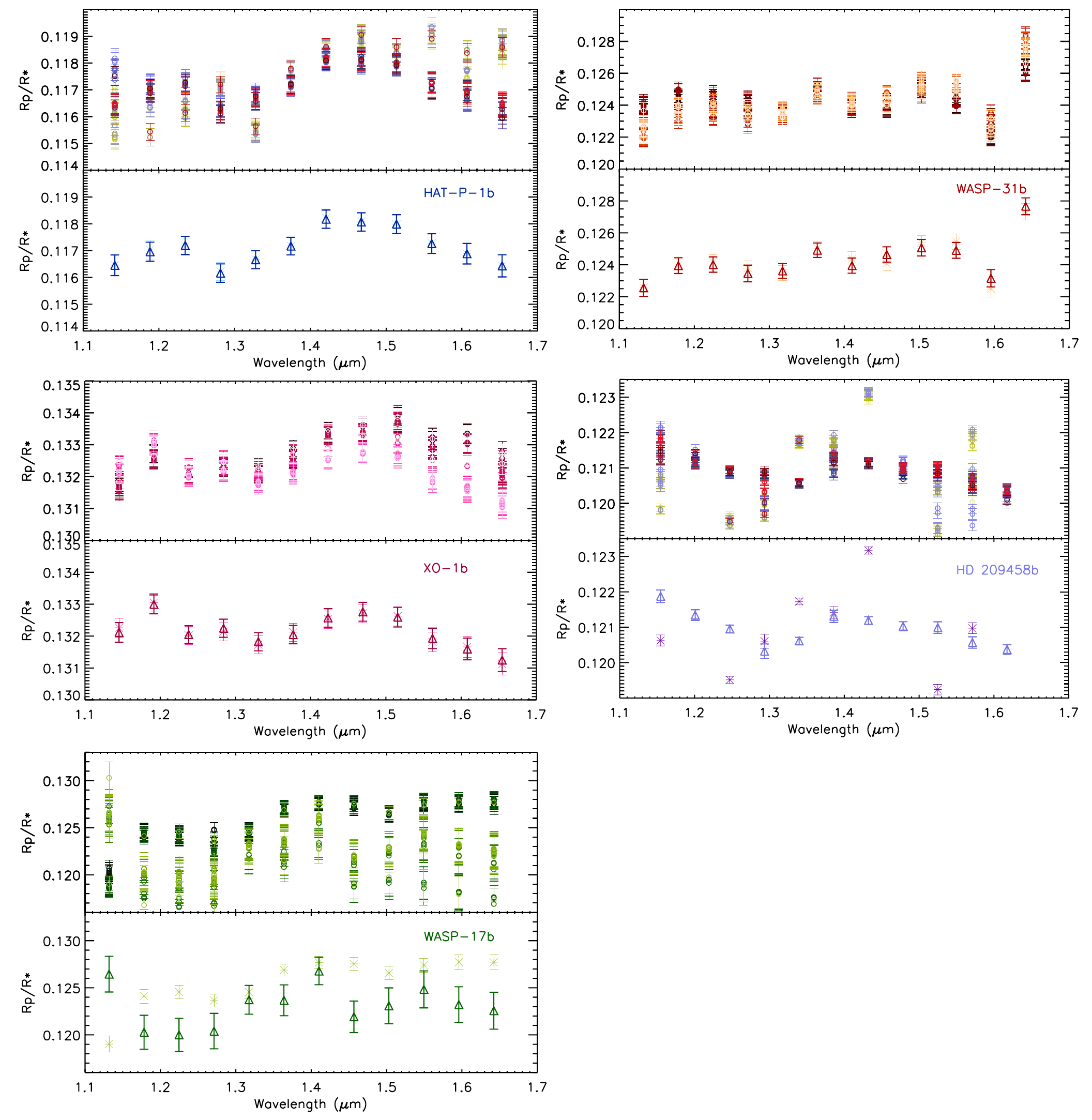}
\caption[\quad Marginalised transmission spectra with spectroscopic wavelength shifts]{The marginalised transmission spectrum for each of our five exoplanet atmospheres using spectroscopic wavelength shifts, $\delta_\lambda(\lambda)$. Top of each plot: the transmission spectrum for each planet calculated using our systematic grid. Bottom of each plot: the marginalised transmission spectrum calculated using systematic wavelength shifts, $\delta_\lambda(\lambda)$ (triangles), and the transmission spectrum calculated using spectroscopic wavelength shifts and the best-fitting systematic model as determined from the band-integrated lightcurve analysis (stars). This shows that in most cases there is little significant difference between the two choices of treatment to the wavelength shifts. However, when the difference between shifts in each spectroscopic bin is large, as seen in HD\,209458b, the spectroscopic treatment of the systematics using the best fitting model from the band-integrated fit is not effective.}
\label{fig:transmission_marginalise_sh}
\vspace{10pt}
\end{figure*}

In Fig. \ref{fig:transmission_marginalise_sh} we show the transmission spectra of all five targets for each systematic model using the $\delta_\lambda(\lambda)$ arrays. Similar to the analysis conducted using the $\delta_\lambda$ arrays, the shape of each transmission spectrum appears robust across most systematic models with variations in the scatter across each of the datasets. In the bottom of each plot we display the marginalised transmission spectrum (triangles) calculated using the whole systematic grid. This demonstrates  how each of the systematic models effect the final R$_p$/R$_*$ value in each bin. Larger scatter represents larger uncertainties on the final points, and where small uncertainties are observed, it is likely the highest weighted models lie in a specific portion of the systematic grid for each bin. In this analysis we also show the transmission spectrum calculated using the highest weighted systematic model, defined by the band-integrated lightcurve in section 4.1 (asterix). In three out of five cases this matches very closely to the marginalised transmission spectrum, suggesting in these cases that additional wavelength dependent systematics do not greatly affect the spectroscopic lightcurves. However, in the case of WASP-17b the transmission spectrum calculated using a single systematic model results in a significant difference in the the shape and uncertainties associated with the spectrum. As seen in the band-integrated lightcurve analysis (Fig. \ref{fig:AIC_weight_all}), this is likely due the small final weighting assigned to any particular systematic model from the grid, so no particular model is adequately able to define the systematics in the lightcurves.

In contrast to the small changes observed between the computed transmission spectra of WASP-17, the transmission spectra computed for HD\,209458b show distinct differences between the two spectra, with increased scatter in over 20\% of the spectral bins and a number of points falling outside the range of the plot. In the case of HD\,209458b's transmission spectrum, using the most favoured band-integrated systematic model to fit the spectroscopic data clearly no-longer applies to the individual bins when using the $\delta_\lambda(\lambda)$ array. This further emphasises the power of the marginalisation technique where the systematics associated with the observations are not well known over a single systematic model fit based on general lightcurve statistics.

\begin{table}
\centering
\caption[\quad HAT-P-1b transmission spectrum]{Table of marginalised $\delta_\lambda(\lambda)$ transmission spectral properties of \mbox{HAT-P-1b} binned to 46.5nm.}
\begin{tabular}{ccc}
\hline
\hline
Wavelength & Rp/R* & $\sigma$ \\ 
\hline
band-integrated & 0.118520 & 0.000730 \\
\hline
1.1416 & 0.116748 & 0.000520 \\ 
1.1881 & 0.116959 & 0.000356 \\ 
1.2346 & 0.117194 & 0.000336 \\ 
1.2811 & 0.116162 & 0.000342 \\ 
1.3276 & 0.116660 & 0.000334 \\ 
1.3741 & 0.117171 & 0.000327 \\ 
1.4206 & 0.118171 & 0.000341 \\ 
1.4671 & 0.118070 & 0.000342 \\ 
1.5136 & 0.117989 & 0.000350 \\ 
1.5601 & 0.117262 & 0.000366 \\ 
1.6066 & 0.116885 & 0.000384 \\ 
1.6531 & 0.116434 & 0.000414 \\ 
\hline
\end{tabular}
\label{table:H1_transmission}
\end{table}

\begin{table}
\centering
\caption[\quad WASP-31b transmission spectrum]{Table of marginalised $\delta_\lambda(\lambda)$ transmission spectral properties of \mbox{WASP-31b} binned to 46.3nm.}
\begin{tabular}{ccc}
\hline
\hline
Wavelength & Rp/R* & $\sigma$ \\
\hline
band-integrated & 0.125400 & 0.000670 \\
\hline
1.1324 & 0.122559 & 0.000534 \\   
1.1787 & 0.123949 & 0.000495 \\   
1.2250 & 0.124015 & 0.000481 \\  
1.2713 & 0.123456 & 0.000523 \\   
1.3176 & 0.123617 & 0.000461 \\    
1.3639 & 0.124915 & 0.000451 \\    
1.4102 & 0.123952 & 0.000479 \\   
1.4565 & 0.124648 & 0.000479 \\    
1.5028 & 0.125068 & 0.000517 \\    
1.5491 & 0.124905 & 0.000496 \\      
1.5954 & 0.123157 & 0.000540 \\   
1.6417 & 0.127668 & 0.000529 \\                
\hline
\end{tabular}
\label{table:W31_transmission}
\end{table}

\begin{table}
\centering
\caption[\quad XO-1b transmission spectrum]{Table of marginalised $\delta_\lambda(\lambda)$ transmission spectral properties of \mbox{XO-1b} binned to 46.3nm.}
\begin{tabular}{ccc}
\hline
\hline
Wavelength & Rp/R* & $\sigma$ \\
\hline
band-integrated & 0.132151 & 0.000401 \\
\hline
1.1452 & 0.132115 & 0.000307 \\      
1.1915 & 0.132994 & 0.000296 \\     
1.2378 & 0.132048 & 0.000276 \\       
1.2841 & 0.132245 & 0.000283 \\    
1.3304 & 0.131824 & 0.000285 \\    
1.3767 & 0.132045 & 0.000289 \\    
1.4230 & 0.132569 & 0.000288 \\      
1.4693 & 0.132762 & 0.000295 \\   
1.5156 & 0.132593 & 0.000307 \\    
1.5619 & 0.131929 & 0.000320 \\   
1.6082 & 0.131596 & 0.000341 \\    
1.6545 & 0.131248 & 0.000356 \\                        
\hline
\end{tabular}
\label{table:XO1_transmission}
\end{table}

\begin{table}
\centering
\caption[\quad WASP-17b transmission spectrum]{Table of marginalised $\delta_\lambda(\lambda)$ transmission spectral properties of \mbox{WASP-17b} binned to 46.4nm.}
\begin{tabular}{ccc}
\hline
\hline
Wavelength & Rp/R* & $\sigma$ \\
\hline
band-integrated & 0.122931 & 0.001270 \\
\hline
1.1318 & 0.125451 & 0.001899 \\      
1.1782 & 0.120278 & 0.001802 \\      
1.2246 & 0.119998 & 0.001758 \\     
1.2710 & 0.120399 & 0.001885 \\    
1.3174 & 0.123727 & 0.001530 \\    
1.3638 & 0.123671 & 0.001641 \\    
1.4102 & 0.126780 & 0.001456 \\    
1.4566 & 0.121919 & 0.001681 \\     
1.5030 & 0.123084 & 0.001905 \\    
1.5494 & 0.124827 & 0.001967 \\    
1.5958 & 0.123212 & 0.001887 \\    
1.6422 & 0.122568 & 0.001965 \\ 
\hline
\end{tabular}
\label{table:W17_transmission}
\end{table}

\begin{table}
\centering
\caption[\quad HD 209458b transmission spectrum]{Table of marginalised `stretch and shift' ($\delta_\lambda(\lambda).A(\lambda)$) transmission spectral properties of \mbox{HD 209458b} binned to 46.5nm.}
\begin{tabular}{ccc}
\hline
\hline
Wavelength & Rp/R* & $\sigma$ \\
\hline
band-integrated & 0.120610 & 0.000291 \\
\hline
1.1683 & 0.120086 & 0.000158 \\    
1.2100 & 0.120226 & 0.000113 \\    
1.2517 & 0.120924 & 0.000120 \\    
1.3934 & 0.120494 & 0.000166 \\     
1.3335 & 0.120456 & 0.000319 \\
1.3767 & 0.121282 & 0.000186 \\     
1.4184 & 0.121234 & 0.000139 \\    
1.4600 & 0.120802 & 0.000098 \\    
1.5017 & 0.120656 & 0.000108 \\    
1.5434 & 0.121144 & 0.000153 \\
\hline
\vspace{10pt}
\end{tabular}
\label{table:HD209_transmission}
\end{table}

\begin{figure*}
\centering
\includegraphics[width=0.95\textwidth]{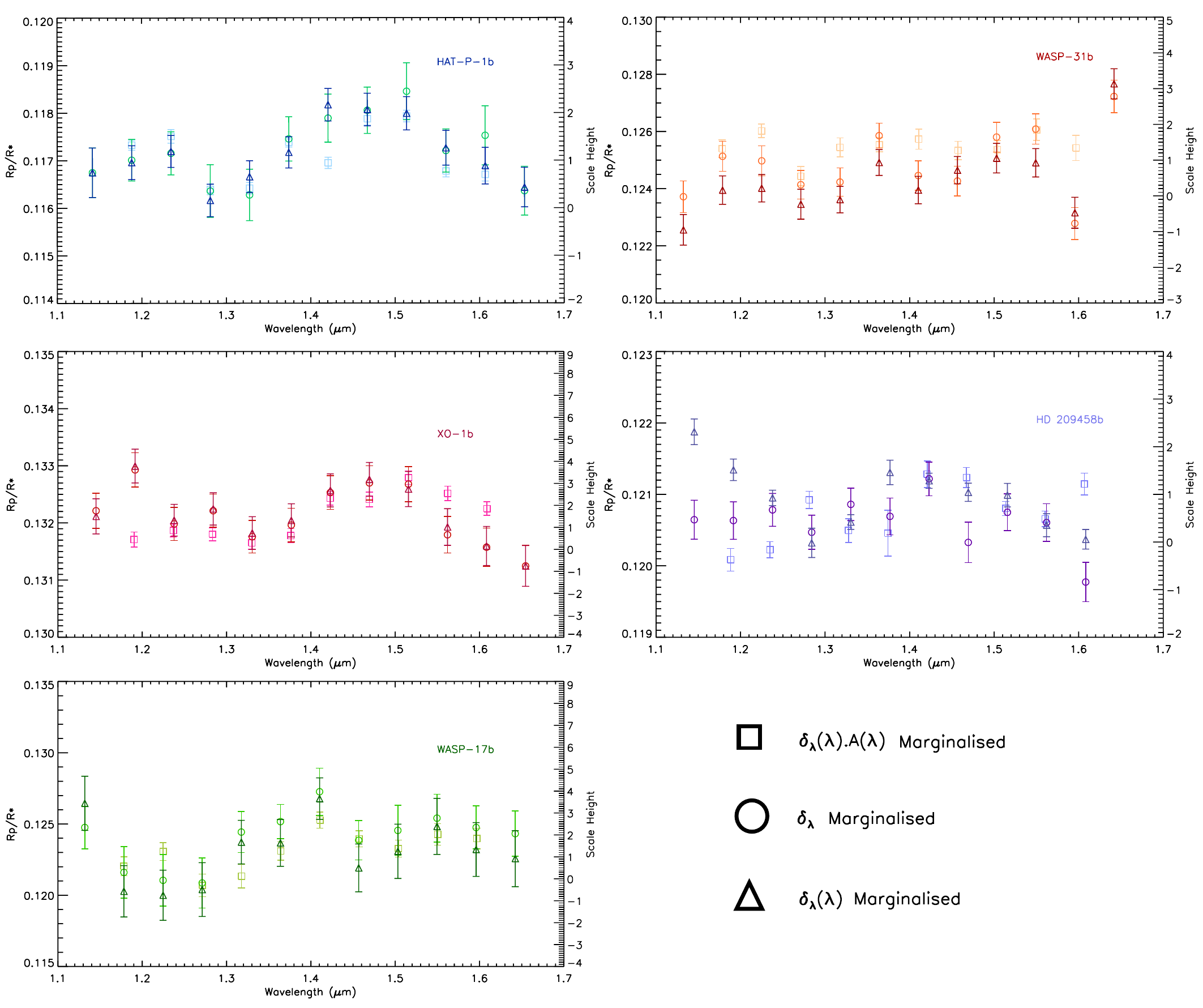}
\caption[\quad Marginalised transmission spectra with stretch and shift]{The marginalised transmission spectrum calculated using stretch and shift $\delta_\lambda(\lambda).A(\lambda)$ (squares), the marginalised transmission spectrum calculated using spectroscopic wavelength shifts $\delta_\lambda(\lambda)$ (triangles), and the marginalised transmission spectrum calculated using the band-integrated wavelength shift $\delta_\lambda$ (circles). Here we present the $\delta_\lambda(\lambda)$ transmission spectrum as the most robust and consistent systematic treatment of WFC3 datasets, to enable large cross-comparisons between multiple observations, and therefore atmospheric detections, when related to the specific planetary scale height (right y-axis).}
\label{fig:marginalised_comparison_methods}
\end{figure*}

\subsubsection{Wavelength stretch and shift template}
To further test the impact of wavelength shift treatment on the resultant transmission spectrum we  additionally apply the wavelength `stretch and shift' template method employed by \citet{deming2013} as described in section 1.1. In each bin the spectrum is smoothed and fit with a template that is shifted in wavelength position and stretched in amplitude ($\delta_\lambda(\lambda).A(\lambda)$), as an averaged spectrum will likely be slightly broadened by the wavelength shifts. This produces a residual array for each wavelength channel with both common-mode band-integrated and wavelength shift systematic corrections removed. 

This method, similar to divide-oot, relies on canceling out common-mode systematics and assumes that HST breathing effects and linear trends in time are both wavelength independent systematics. To account for any residual wavelength dependent systematics in the spectroscopic lightcurves, we modify the `stretch and shift' technique to run with our systematic grid to determine if any additional systematics are present in the spectroscopic lightcurve. We then marginalise over the whole systematic grid to calculate the final transmission spectrum for each planet.

\subsection{Marginalised transmission spectra}
In Fig. \ref{fig:marginalised_comparison_methods} we show the margainalised transmission spectrum from all three of our wavelength shift treatments; $\delta_\lambda$ (circles), $\delta_\lambda(\lambda)$ (triangles), and $\delta_\lambda(\lambda).A(\lambda)$ (squares). The transmission spectra for HAT-P-1b, WASP-31b, XO-1b, and WASP-17b, show no statistically significant differences between each of the three treatments. Importantly, this demonstrates that the more intensive treatments to the wavelength shift systematics are not producing additional non-physical absorption signatures in the transmission spectrum. 

The transmission spectrum of HD\,209458b, however, shows large differences between all three of the methods employed to correct the significant wavelength shifts that occurred over the course of the observations. This strongly suggests that the shifts in wavelength observed at the band-integrated level, and at the spectroscopic level, have a large effect on the systematic treatment and therefore interpretation that can be placed on the calculated transmission spectrum. Using the $\delta_\lambda$ correction produces a flat transmission spectrum with large uncertainties, while structure emerges with reduced uncertainties using both the $\delta_\lambda(\lambda)$ and the `stretch and shift' $\delta_\lambda(\lambda).A(\lambda)$ techniques. This suggests that there are wavelength dependent systematics associated with the wavelength shifts in the spectroscopic lightcurves which are used to calculate the atmospheric transmission spectrum. 

In this study we present marginalisation as a method of systematic correction which can be applied to a large series of datasets simultaneously using the individual datasets to define the corrections applied. As such, the systematic models being mraginalised over need to represent the diversity in datasets and observation strategies. Here we present marginalisation along with the $\delta_\lambda(\lambda)$ treatment as a main result of our analysis, as it treats the wavelength dependence of the observational systematics in a consistent manner as defined by the individual dataset, while still allowing for a direct comparison between each planetary atmosphere, by maintaining the relative  absolute depth of the observed transit. The $\delta_\lambda(\lambda)$ correction is favoured over the stretch and shift method, as stretch and shift uses smoothing and common-mode techniques which adjust the absolute baseline of the transmission spectrum making interpretation of low amplitude spectra difficult. The $\delta_\lambda(\lambda)$ correction is also favoured over the broad-band $\delta_\lambda$ correction as it accounts for the small variations in the wavelength shifts measured in each bin which impact the data rather than using the average, as the deviation from the average can vary from observation to observation (see Fig. \ref{fig:wavelength_shifts}). We quote the computed transmission spectra for each of the planetary datasets following marginalisation of our systematic grid with the $\delta_\lambda(\lambda)$ correction in Tables \ref{table:H1_transmission} to \ref{table:W17_transmission}. In the case of HD\,209458 where fast scan rates were used as a result of a bright target star, we present the `stretch and shift' method, $\delta_\lambda(\lambda.A(\lambda)$, as brighter targets require an additional level of processing to produce robust and accurate transmission spectra (Table \ref{table:HD209_transmission}).

\begin{figure}
\centering 
\includegraphics[width=0.47\textwidth]{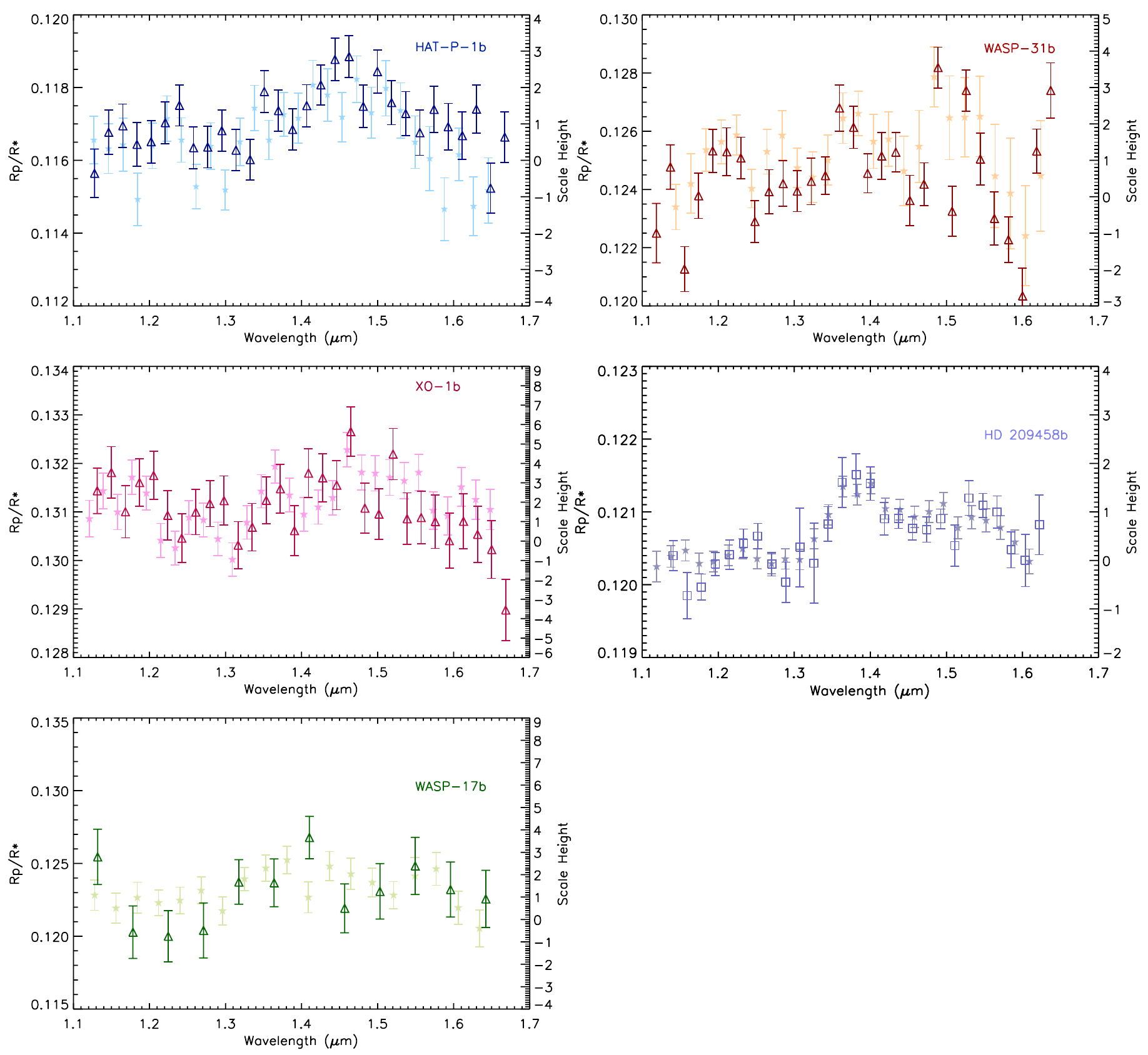}
\caption[\quad Marginalised transmission spectra vs published spectra]{Marginalised transmission spectrum calculated using spectroscopic wavelength shifts $\delta_\lambda(\lambda)$ (triangles) plotted against the published transmission spectra (stars). HAT-P-1b (\citealt{wakeford2013}), WASP-31b (\citealt{Sing2015}), XO-1b, and WASP-17b (\citealt{mandell2013}), and the transmission spectrum computed using `stretch and shift' (squares) for HD\,209458b (\citealt{deming2013})}
\label{fig:published}
\end{figure}

Figure \ref{fig:published} shows the marginalised transmission spectra computed for all of our targets compared to the previously published transmission spectral values. Where appropriate we have attempted to match the resolution of the previous measurements, in the case of WASP-17b this was not possible due to the quality of the data and the lower resolution spectrum is shown. In the case of XO-1b and HD\,209458b the published transmission spectra have been shifted by -0.0016 and -0.00022, respectively as the published spectra include a common-mode technique which affects the absolute baseline of the calculated transmission spectra. This shows that the marginalisation technique can be used both on the band-integrated lightcurve, and the spectroscopic lightcurves while maintaining the absolute depth and shape of transit. For HD\,209458b we find larger uncertainties associated with several bins as marginalisation incorporates the information from all models in the systematic grid, while the published data is fit only with a linear trend in time.

%
\section{Conclusion}

We have applied the systematic marginalisation technique proposed by \citet{gibson2014} to HST WFC3 lightcurves to determine robust transit parameters across multiple datasets. This has allowed us to incorporate all of our current knowledge of HST WFC3 into a single systematic treatment when estimating the planet parameters. We examine the marginalisation technique in relation to the current field and present an outline of the methods used. We compute the maximum likelihood estimation, for each systematic model when fit to the data based on the AIC. We then calculate the weight assigned to each systematic model and use the information from all tested models to calculate the final marginalised transit parameters for both the band-integrated lightcurve, and the spectroscopic lightcurves to construct the atmospheric transmission spectrum. This allows for a more through exploration of the degeneracies between the planet signal and the affecting systematics to derive more realistic uncertainties on the data.  

We demonstrate that marginalisation is able to more accurately describe the data producing realistic uncertainties, as well as providing valuable insight into the instrument and impacting systematics. We find that the systematic models with the heighest weight all include a higher order (3rd or 4th order) polynomial correction in HST phase, to account for the thermal variations over the course of each orbit. We also find that a linear correction in time across the whole visit is required in over five out of the top ten systematic models selected for each dataset, suggesting it has a high impact on the observational systematics in WFC3 data. We additionally test the dependence on the shift in spectral wavelength position over the course of the observations and find that spectroscopic wavelength shifts $\delta_\lambda(\lambda)$, best describe the associated systematic when applying the full systematic grid to each spectroscopic lightcurve to produce the final transmission spectrum. However, in the case of bright targets where scan rates are large, such as HD\,209458b, additional treatment to the spectral position systematic are required to produce a robust and accurate transmission spectrum.

The use of marginalisation accounts for all of the information from each systematic model tested, compared to relying solely on the BIC to select a particular systematic model for the data. Marginalisation across a systematic grid is shown to have the greatest effect on datasets where the impacting systematics are not well defined but can be estimated by a family of similar models. Using marginalisation to correct for instrument systematics in both the band-integrated and spectroscopic lightcurves, we show the measured transmission spectrum of five exoplanet atmospheres which maintain the absolute transit depth. In most cases, we find similar overall shapes and baseline depths using marginalisation. Compared to previously published results, marginalisation can increases the errorbar values on each measurement by up to $\sim$30ppm in the most extreme cases, when the systematics cannot be well described by known systematics.

Critically, we contrast this technique with current analysis methods used in the field and demonstrate the uses of marginalisation to make large comparative studies of different exoplanet atmospheres possible. Previous studies often use the most favoured systematic model alone to correct transit lightcurves, we show that using marginalisation across a systematic grid of models can more accurately represent data where multiple systematic models fit equally well. The use of marginalisation over a grid of systematic models related to the instrument where observations strategy and instrument modes may have a varied effect, allows for a direct comparison between multiple planetary atmospheres by allowing the data to define the analysis while keeping the characterisation consistent, which cannot be done when varied singular analysis methods are applied. 

Marginalisation can be applied to a multitude of transit observations, expanding the ability to make a robust comparison of exoplanet atmospheres as more favourable targets are observed. Additionally, margianalising over a series of systematic models will be important for new generations of instrument, to more easily determine the impact individual systematics have on the data and to aid in optimised observing strategies. This will be particularly important for James Webb Space Telescope observations where there is potential to observe smaller and colder worlds, as the observing strategies and treatment of systematics will likely define the measurements. 

As can be seen in Fig. \ref{fig:marginalised_comparison_methods} the transmission spectra show an overall similarity around the H$_2$O absorption band centered at 1.4\,$\mu$m with varied depressed amplitude features. In a follow-up paper we use marginalisation across a wider range of hot Jupiter transit lightcurves from WFC3 in a comparative study of their atmospheres and discuss the interpretation with a series of theoretical models and T-P profiles. 

%
\section{Acknowledgements}
The authors would like to thank N. Gibson for useful comments and discussions on this paper and the analysis technique presented. H.R. Wakeford acknowledges support by an appointment to the NASA Postdoctoral Program at Goddard Space Flight Center, administered by Oak Ridge Associated Universities through a contract with NASA. H.R. Wakeford, D.K. Sing, and T. Evans acknowledge funding from the European Research Council under the European Union’s Seventh Framework Programme (FP7/2007-2013) / ERC grant agreement no. 336792.
This work is based on observations with the NASA/ESA Hubble Space Telescope. This research has made use of NASAs Astrophysics Data System, and components of the IDL astronomy library. \\



{\it Facilities:} \facility{HST (WFC3)}. \\
%
\bibliographystyle{mn2e}
\bibliography{references}

\begin{thebibliography}{}

\bibitem[\protect\citeauthoryear{Berta, Charbonneau, D{\'e}sert, Kempton,
  McCullough, Burke, Fortney, Irwin, Nutzman \& Homeier}{Berta
  et~al.}{2012}]{berta2012}
Berta Z.~K.,  Charbonneau D.,  D{\'e}sert J.-M.,  Kempton E. M.-R.,  McCullough
  P.~R.,  Burke C.~J.,  Fortney J.~J.,  Irwin J.,  Nutzman P.,    Homeier D.,
  2012, \apj, 747, 35

\bibitem[\protect\citeauthoryear{{Brown}}{{Brown}}{2001}]{brown2001}
{Brown} T.~M.,  2001, \apj, 553, 1006

\bibitem[\protect\citeauthoryear{Burnham \& Anderson}{Burnham \&
  Anderson}{2004}]{burnham2004}
Burnham K.~P.,  Anderson D.~R.,  2004, Sociological methods \& research, 33,
  261

\bibitem[\protect\citeauthoryear{{Claret}}{{Claret}}{2000}]{claret2000}
{Claret} A.,  2000, \aap, 363, 1081

\bibitem[\protect\citeauthoryear{{Crossfield}, {Barman}, {Hansen} \&
  {Howard}}{{Crossfield} et~al.}{2013}]{crossfield2013}
{Crossfield} I.~J.~M.,  {Barman} T.,  {Hansen} B.~M.~S.,    {Howard} A.~W.,
  2013, \aap, 559, A33

\bibitem[\protect\citeauthoryear{Deming, Wilkins, McCullough, Burrows, Fortney,
  Agol, Dobbs-Dixon, Madhusudhan, Crouzet, Desert et~al.,}{Deming
  et~al.}{2013}]{deming2013}
Deming D.,  Wilkins A.,  McCullough P.,  Burrows A.,  Fortney J.~J.,  Agol E.,
  Dobbs-Dixon I.,  Madhusudhan N.,  Crouzet N.,  Desert J.-M.,    et~al., 2013,
  \apj, 774, 95

\bibitem[\protect\citeauthoryear{{Ehrenreich}, {Bonfils}, {Lovis}, {Delfosse},
  {Forveille}, {Mayor}, {Neves}, {Santos}, {Udry} \&
  {S{\'e}gransan}}{{Ehrenreich} et~al.}{2014}]{ehrenreich2014}
{Ehrenreich} D.,  {Bonfils} X.,  {Lovis} C.,  {Delfosse} X.,  {Forveille} T.,
  {Mayor} M.,  {Neves} V.,  {Santos} N.~C.,  {Udry} S.,    {S{\'e}gransan} D.,
  2014, \aap, 570, A89

\bibitem[\protect\citeauthoryear{{Fraine}, {Deming}, {Benneke}, {Knutson},
  {Jord{\'a}n}, {Espinoza}, {Madhusudhan}, {Wilkins} \& {Todorov}}{{Fraine}
  et~al.}{2014}]{fraine2014}
{Fraine} J.,  {Deming} D.,  {Benneke} B.,  {Knutson} H.,  {Jord{\'a}n} A.,
  {Espinoza} N.,  {Madhusudhan} N.,  {Wilkins} A.,    {Todorov} K.,  2014,
  \nat, 513, 526

\bibitem[\protect\citeauthoryear{{Gibson}}{{Gibson}}{2014}]{gibson2014}
{Gibson} N.~P.,  2014, \mnras, 445, 3401

\bibitem[\protect\citeauthoryear{{Gibson}, {Aigrain}, {Pont}, {Sing},
  {D{\'e}sert}, {Evans}, {Henry}, {Husnoo} \& {Knutson}}{{Gibson}
  et~al.}{2012}]{gibson2012b}
{Gibson} N.~P.,  {Aigrain} S.,  {Pont} F.,  {Sing} D.~K.,  {D{\'e}sert} J.-M.,
  {Evans} T.~M.,  {Henry} G.,  {Husnoo} N.,    {Knutson} H.,  2012, \mnras,
  422, 753

\bibitem[\protect\citeauthoryear{{Haynes}, {Mandell}, {Madhusudhan}, {Deming}
  \& {Knutson}}{{Haynes} et~al.}{2015}]{haynes2015}
{Haynes} K.,  {Mandell} A.~M.,  {Madhusudhan} N.,  {Deming} D.,    {Knutson}
  H.,  2015, \apj, 806, 146

\bibitem[\protect\citeauthoryear{{Huitson}, {Sing}, {Pont}, {Fortney},
  {Burrows}, {Wilson}, {Ballester}, {Nikolov}, {Gibson}, {Deming}, {Aigrain},
  {Evans}, {Henry}, {Lecavelier des Etangs}, {Showman}, {Vidal-Madjar} \&
  {Zahnle}}{{Huitson} et~al.}{2013}]{huitson2013}
{Huitson} C.~M.,  {Sing} D.~K.,  {Pont} F.,  {Fortney} J.~J.,  {Burrows} A.~S.,
   {Wilson} P.~A.,  {Ballester} G.~E.,  {Nikolov} N.,  {Gibson} N.~P.,
  {Deming} D.,  {Aigrain} S.,  {Evans} T.~M.,  {Henry} G.~W.,  {Lecavelier des
  Etangs} A.,  {Showman} A.~P.,  {Vidal-Madjar} A.,    {Zahnle} K.,  2013,
  \mnras, 434, 3252

\bibitem[\protect\citeauthoryear{{Knutson}, {Dragomir}, {Kreidberg}, {Kempton},
  {McCullough}, {Fortney}, {Bean}, {Gillon}, {Homeier} \& {Howard}}{{Knutson}
  et~al.}{2014}]{knutson2014}
{Knutson} H.~A.,  {Dragomir} D.,  {Kreidberg} L.,  {Kempton} E.~M.-R.,
  {McCullough} P.~R.,  {Fortney} J.~J.,  {Bean} J.~L.,  {Gillon} M.,  {Homeier}
  D.,    {Howard} A.~W.,  2014, \apj, 794, 155

\bibitem[\protect\citeauthoryear{{Kreidberg}, {Bean}, {D{\'e}sert}, {Benneke},
  {Deming}, {Stevenson}, {Seager}, {Berta-Thompson}, {Seifahrt} \&
  {Homeier}}{{Kreidberg} et~al.}{2014}]{kreidberg2014a}
{Kreidberg} L.,  {Bean} J.~L.,  {D{\'e}sert} J.-M.,  {Benneke} B.,  {Deming}
  D.,  {Stevenson} K.~B.,  {Seager} S.,  {Berta-Thompson} Z.,  {Seifahrt} A.,
   {Homeier} D.,  2014, \nat, 505, 69

\bibitem[\protect\citeauthoryear{{Kreidberg}, {Bean}, {D{\'e}sert}, {Line},
  {Fortney}, {Madhusudhan}, {Stevenson}, {Showman}, {Charbonneau},
  {McCullough}, {Seager}, {Burrows}, {Henry}, {Williamson}, {Kataria} \&
  {Homeier}}{{Kreidberg} et~al.}{2014}]{kreidberg2014b}
{Kreidberg} L.,  {Bean} J.~L.,  {D{\'e}sert} J.-M.,  {Line} M.~R.,  {Fortney}
  J.~J.,  {Madhusudhan} N.,  {Stevenson} K.~B.,  {Showman} A.~P.,
  {Charbonneau} D.,  {McCullough} P.~R.,  {Seager} S.,  {Burrows} A.,  {Henry}
  G.~W.,  {Williamson} M.,  {Kataria} T.,    {Homeier} D.,  2014, \apjl, 793,
  L27

\bibitem[\protect\citeauthoryear{{Line}, {Knutson}, {Deming}, {Wilkins} \&
  {Desert}}{{Line} et~al.}{2013}]{line2013}
{Line} M.~R.,  {Knutson} H.,  {Deming} D.,  {Wilkins} A.,    {Desert} J.-M.,
  2013, \apj, 778, 183

\bibitem[\protect\citeauthoryear{{Mandel} \& {Agol}}{{Mandel} \&
  {Agol}}{2002}]{mandelagol2002}
{Mandel} K.,  {Agol} E.,  2002, \apjl, 580, L171

\bibitem[\protect\citeauthoryear{{Mandell}, {Haynes}, {Sinukoff},
  {Madhusudhan}, {Burrows} \& {Deming}}{{Mandell} et~al.}{2013}]{mandell2013}
{Mandell} A.~M.,  {Haynes} K.,  {Sinukoff} E.,  {Madhusudhan} N.,  {Burrows}
  A.,    {Deming} D.,  2013, \apj, 779, 128

\bibitem[\protect\citeauthoryear{{Markwardt}}{{Markwardt}}{2009}]{markwardt2009}
{Markwardt} C.~B.,  2009, in {Bohlender} D.~A.,  {Durand} D.,   {Dowler} P.,
  eds, Astronomical Data Analysis Software and Systems XVIII Vol.~411 of
  Astronomical Society of the Pacific Conference Series, {Non-linear
  Least-squares Fitting in IDL with MPFIT}.
p.~251

\bibitem[\protect\citeauthoryear{{McCullough}}{{McCullough}}{2011}]{mccullough2011}
{McCullough} P.,  2011, WFC Space Telescope Analysis Newsletter 6

\bibitem[\protect\citeauthoryear{{McCullough}, {Crouzet}, {Deming} \&
  {Madhusudhan}}{{McCullough} et~al.}{2014}]{mccullough2014}
{McCullough} P.~R.,  {Crouzet} N.,  {Deming} D.,    {Madhusudhan} N.,  2014,
  \apj, 791, 55

\bibitem[\protect\citeauthoryear{{Nikolov}, {Sing}, {Pont}, {Burrows},
  {Fortney}, {Ballester}, {Evans}, {Huitson}, {Wakeford}, {Wilson} \& {et
  al.}}{{Nikolov} et~al.}{2014}]{nikolov2014}
{Nikolov} N.,  {Sing} D.~K.,  {Pont} F.,  {Burrows} A.~S.,  {Fortney} J.~J.,
  {Ballester} G.~E.,  {Evans} T.~M.,  {Huitson} C.~M.,  {Wakeford} H.~R.,
  {Wilson} P.~A.,    {et al.} 2014, \mnras, 437, 46

\bibitem[\protect\citeauthoryear{{Ranjan}, {Charbonneau}, {D{\'e}sert},
  {Madhusudhan}, {Deming}, {Wilkins} \& {Mandell}}{{Ranjan}
  et~al.}{2014}]{ranjan2014}
{Ranjan} S.,  {Charbonneau} D.,  {D{\'e}sert} J.-M.,  {Madhusudhan} N.,
  {Deming} D.,  {Wilkins} A.,    {Mandell} A.~M.,  2014, \apj, 785, 148

\bibitem[\protect\citeauthoryear{Sing, Fortney, Nikolov, Wakeford, Kataria,
  Evans, Aigrain, Ballester, Burrows, Deming, D{\'e}sert, Gibson, Henry,
  Huitson, Knutson, Etangs, Pont, Showman, Vidal-Madjar, Williamson \&
  Wilson}{Sing et~al.}{2016}]{sing2016}
Sing D.~K.,  Fortney J.~J.,  Nikolov N.,  Wakeford H.~R.,  Kataria T.,  Evans
  T.~M.,  Aigrain S.,  Ballester G.~E.,  Burrows A.~S.,  Deming D.,  D{\'e}sert
  J.-M.,  Gibson N.~P.,  Henry G.~W.,  Huitson C.~M.,  Knutson H.~A.,  Etangs
  A. L.~d.,  Pont F.,  Showman A.~P.,  Vidal-Madjar A.,  Williamson M.~H.,
  Wilson P.~A.,  2016, Nature, 529, 59

\bibitem[\protect\citeauthoryear{{Sing}, {Lecavelier des Etangs}, {Fortney},
  {Burrows}, {Pont}, {Wakeford}, {Ballester}, {Nikolov}, {Henry} \& {et
  al.}}{{Sing} et~al.}{2013}]{sing2013}
{Sing} D.~K.,  {Lecavelier des Etangs} A.,  {Fortney} J.~J.,  {Burrows} A.~S.,
  {Pont} F.,  {Wakeford} H.~R.,  {Ballester} G.~E.,  {Nikolov} N.,  {Henry}
  G.~W.,    {et al.} 2013, \mnras, 436, 2956

\bibitem[\protect\citeauthoryear{{Stevenson}, {Bean}, {Fabrycky} \&
  {Kreidberg}}{{Stevenson} et~al.}{2014}]{stevenson2014c}
{Stevenson} K.~B.,  {Bean} J.~L.,  {Fabrycky} D.,    {Kreidberg} L.,  2014,
  \apj, 796, 32

\bibitem[\protect\citeauthoryear{{Stevenson}, {Bean}, {Seifahrt}, {D{\'e}sert},
  {Madhusudhan}, {Bergmann}, {Kreidberg} \& {Homeier}}{{Stevenson}
  et~al.}{2014}]{Stevenson2014a}
{Stevenson} K.~B.,  {Bean} J.~L.,  {Seifahrt} A.,  {D{\'e}sert} J.-M.,
  {Madhusudhan} N.,  {Bergmann} M.,  {Kreidberg} L.,    {Homeier} D.,  2014,
  \aj, 147, 161

\bibitem[\protect\citeauthoryear{{Stevenson}, {D{\'e}sert}, {Line}, {Bean},
  {Fortney}, {Showman}, {Kataria}, {Kreidberg}, {McCullough}, {Henry},
  {Charbonneau}, {Burrows}, {Seager}, {Madhusudhan}, {Williamson} \&
  {Homeier}}{{Stevenson} et~al.}{2014}]{stevenson2014b}
{Stevenson} K.~B.,  {D{\'e}sert} J.-M.,  {Line} M.~R.,  {Bean} J.~L.,
  {Fortney} J.~J.,  {Showman} A.~P.,  {Kataria} T.,  {Kreidberg} L.,
  {McCullough} P.~R.,  {Henry} G.~W.,  {Charbonneau} D.,  {Burrows} A.,
  {Seager} S.,  {Madhusudhan} N.,  {Williamson} M.~H.,    {Homeier} D.,  2014,
  Science, 346, 838

\bibitem[\protect\citeauthoryear{{Swain}, {Deroo}, {Tinetti}, {Hollis},
  {Tessenyi}, {Line}, {Kawahara}, {Fujii}, {Showman} \& {Yurchenko}}{{Swain}
  et~al.}{2013}]{swain2013}
{Swain} M.,  {Deroo} P.,  {Tinetti} G.,  {Hollis} M.,  {Tessenyi} M.,  {Line}
  M.,  {Kawahara} H.,  {Fujii} Y.,  {Showman} A.~P.,    {Yurchenko} S.~N.,
  2013, \icarus, 225, 432

\bibitem[\protect\citeauthoryear{Wakeford, Sing, Deming, Gibson, Fortney,
  Burrows, Ballester, Nikolov, Aigrain, Henry et~al.,}{Wakeford
  et~al.}{2013}]{wakeford2013}
Wakeford H.,  Sing D.,  Deming D.,  Gibson N.,  Fortney J.,  Burrows A.,
  Ballester G.,  Nikolov N.,  Aigrain S.,  Henry G.,    et~al., 2013, \mnras,
  435, 3481

\bibitem[\protect\citeauthoryear{{Wilkins}, {Deming}, {Madhusudhan}, {Burrows},
  {Knutson}, {McCullough} \& {Ranjan}}{{Wilkins} et~al.}{2014}]{wilkins2014}
{Wilkins} A.~N.,  {Deming} D.,  {Madhusudhan} N.,  {Burrows} A.,  {Knutson} H.,
   {McCullough} P.,    {Ranjan} S.,  2014, \apj, 783, 113

\end{thebibliography}

\end{document}